\documentclass{article}

\usepackage[utf8]{inputenc}
\usepackage{geometry}
\usepackage{times}
\usepackage{setspace}
\usepackage{hyperref}
\usepackage{authblk}
\usepackage{graphicx} % Add this package for figures
\usepackage{rotating}
\usepackage{multirow}
\usepackage{natbib}
\usepackage{float}
\usepackage{siunitx}
\usepackage{upgreek}
\newcommand{\teff}{T$_{\rm eff}$}
\newcommand{\logg}{$\log$g}
\newcommand{\kep}{{\it Kepler}}
\newcommand{\gaia}{{\it Gaia}}

\newcommand{\etl}{{\it et al.\,}}

\title{A variable star population in the open cluster NGC\,6819 observed by the \kep\ spacecraft}
\author{\small S.\,Sanjayan$^{1,2}$, A.S.\,Baran$^{1,3}$, P.\,N\'emeth$^{1,4,5}$ and K.\,Kinemuchi$^{6,7}$\\
$^{1}$ARDASTELLA Research Group\\
$^{2}$Centrum Astronomiczne im. Miko{\l}aja Kopernika, Polskiej Akademii Nauk, ul. Bartycka 18, 00-716 Warszawa, Polska\\
$^{3}$Astronomical Observatory, University of Warsaw, Al. Ujazdowskie 4, 00-478 Warszawa, Poland\\
$^{4}$Astroserver.org, F\H{o} t\'er 1, 8533 Malomsok, Hungary\\
$^{5}$Astronomical Institute of the Czech Academy of Sciences, Fri\v{c}ova 298, CZ-251\,65 Ond\v{r}ejov, Czech Republic\\
$^{6}$Department of Astronomy, New Mexico State University, Box 30001, MSC 4500, Las Cruces, NM 88003, USA\\
$^{7}$Apache Point Observatory, 2001 Apache Point Road, P.O. Box 59, Sunspot, NM 88349-0059
}
\date{November 22, 2022}

\begin{document}

\maketitle

\begin{abstract}
We present the list of variable stars we found in the \kep\ superstamp data covering approximately nine arcminutes from the central region of NGC\,6819. This is a continuation of our work presented by Sanjayan \etl(2022a). We classified the variable stars based on the variability type and we established their cluster membership based on the available \gaia\ Data Release\,3 astrometry. Our search revealed 385 variable stars but only 128 were found to be cluster members. In the case of eclipsing and contact binaries we calculated the mid-times of eclipses and derived ephemerides. We searched for eclipse timing variation using the observed minus calculated diagrams. Only five objects show significant orbital period variation. We used isochrones calculated within the MESA Isochrones and Stellar Tracks project and derived the average age (2.54\,Gyr), average distance (2.3\,kpc) and iron content [Fe/H]\,=\,-0.01(2), of NGC\,6819. We confirm this distance by the one derived from \gaia\ astrometry of the cluster members with membership probabilities greater than 0.9.\\
{\bf Key words: NGC\,6819 - Open cluster - Kepler Mission - Variable stars}
\end{abstract}

\section{Introduction}
%About the cluster
In the notes of Caroline Herschel\,(Hoskin\,2005), NGC\,6819 was listed in the class of nebulae and star clusters found between $\theta$\,Lyra and $\delta$\,Cygni. Early study was mostly focused on photometry that allowed for construction of a color\,--\,magnitude diagram\,(CMD) and to estimate the age of the cluster. Detecting only a weak nebula in photographic plates from the Palomar Observatory Sky Survey (POSS-I), King\,(1964) added NGC\,6819 to the list of open clusters that may be old, which made the cluster exceptional as open clusters were considered to be young objects. The first photometry of 38 stars in the field of the cluster were done by Purgathofer\,(1966) who used the original plates from Barnard\,(1931). From a color-magnitude diagram Burkhead\,(1971) pointed out that the main sequence turnoff (MSTO) region of NGC\,6819 is near to that of M67 hence both clusters have comparable ages, however no age of the latter was reported. Later Lindoff\,(1972) and Auner\,(1974) observed the cluster in the B-V color. They compared the MSTO and red giant branch\,(RGB) with other older clusters and predicted that the age of the cluster is at least 2.5\,Gyr. From the theoretical isochrones Rosvick and Vandenberg\,(1998) reported an age of 2.4\,Gyr, which is in agreement with the age 2.3(2)\,Gyr estimated by Anthony-Twarog \etl(2014). Rosvick and Vandenberg\,(1998) suggested that red clump\,(RC) stars are the descendants of blue stragglers\,(BS), given that NGC\,6819 has a confirmed BS population. Kalirai \etl(2001) estimated a total mass of the cluster to be 2\,600\,M$_{\odot}$.

%variability survey literature
Many researchers undertook a search for a variable content in the area of the cluster. We compared our findings with the results reported in most of the following papers. Barkhatova and Vasilevsky\,(1967) conducted the first attempt to detect variable stars in NGC\,6819 with a null result. A variable of the irregular type was reported for the first time by Lindoff\,(1971). As a result of search for contact binaries Kaluzny and Shara\,(1988) made the first significant discovery by detecting 11 variable stars but none of them were reported to be a contact binary. Manteiga \etl(1989) used RV data of stars in the area of the cluster and detected only one RV variable BS. Kryachko\,(2001) discovered two dwarf novae in the cluster area. A survey by Street \etl(2002) revealed 25 variable stars and 13 suspected variable stars in the area of NGC\,6819. Street \etl(2003) did an extended search for planetary transits by analysing data of 38\,000 stars in the area of NGC\,6819, and they found 11 stars with transiting events. Street \etl(2005) reported a detection of 141 variable stars, including 53 eclipsing binaries, eight RS\,CVn variables, 70 rotational stars, 13 long period variables and five variables of other types. Stello \etl(2010) reported the first asteroseismic analysis of solar-like red giants in the cluster. Additional ground-based observations were taken by Talamantes \etl(2010) in the area of NGC\,6819 and they listed 14 binaries and one $\delta$\,Scuti pulsator. Using the XMM\,Newton data, Gosnell \etl(2012) made an unexpected discovery of 12 X-ray sources in the cluster area.

A time-series photometry of the cluster variable population allowed for an independent determination of a distance modulus and cluster age. Balona \etl(2013) used 129 rotational variables in NGC\,6819 to study the relation between the period and B-V color, and they derived cluster age of about 2.5\,Gyr. Basu \etl(2011) estimated the distance modulus of (m-M)$_0\approx$\,11.85\,mag and reddening E(B-V)$\approx$0.15\,mag using asteroseismic analysis of red giant stars in the cluster. The authors estimated the distance to the cluster to be 2\,344(24)\,pc. According to Salaris \etl(2004), NGC\,6819 is located in the halo of the Milky Way and 300\,pc above the galactic plane.

Spectroscopic observations allowed for determination of the motion in the Galaxy and the cluster metallicity. Using only seven cluster members, Friel \etl(1989) derived the mean cluster radial velocity (RV) of -7(13)\,km/s. A more precise estimation of 2.34(5)\,km/s was reported by Hole \etl(2009). Ak\,T \etl(2016) reported that the cluster has a slightly eccentric orbit with the eccentricity\,(\textit{e}) of 0.06 and has the orbital period of 142\,Myr. According to Bragaglia \etl(2001), the cluster has a super solar metallicity of [Fe/H]\,=\,+0.09(3), while Lee \etl(2015) pointed to a sub solar metallicity of [Fe/H]\,=\,-0.02(2).

%Literature search on membership studies conducted on NGC6819
Stellar clusters are studied through their members, hence it is essential to identify cluster members. Below, we present a historical progress of a member hunt in NGC\,6819. Sanders\,(1972) used the relative proper-motion of 189 stars in the cluster area and found 88 probable cluster members. Using RV measurements from the WIYN open cluster study\,XXIV, Hole \etl(2009) established membership of 913 stars in the cluster area, finding 437 targets to have probabilities greater than 50\%. Stello \etl(2011) determined membership probability of 61 stars using asteroseismic analysis and found 50 members. Platais \etl(2013) derived 2\,314 stars with more than 50\% membership probability among 15\,750 stars located in the cluster area. The authors used data from the WIYN open cluster study LV. Gao \etl(2015) established 537 cluster members by means of RVs and proper motion data. Zhang \etl(2015) presented membership studies of 80 stars using RVs from the Large Sky Area Multi-Object Fibre Spectroscopic Telescope (LAMOST) survey. Sampedro \etl(2017) determined the membership probability of 1\,074 stars using the United States Naval Observatory\,(USNO) CCD Astrograph Catalog\,(UCAC4). The first membership analysis using \gaia\ astrometry was done by Cantat-Gaudin \etl(2018), who found 1\,915 members. The cited works revealed many confirmed cluster members derived by means of RVs, proper motions, and seismic analysis. In Section\,4 we present a comparison of the results of the membership studies reported by Cantat-Gaudin \etl(2018) with our results.

To increase the efficiency of the detection of all variable star populations in the cluster area, we undertook an analysis of all \kep\ superstamp data of NGC\,6819. The variable stars should in future serve the purpose of deriving the most precise parameters of the cluster, through binary star, asteroseismic and gyrochronology studies. In addition, \gaia\ data of the cluster members were used toward the cluster age and distance derivation. This is a continuation of our previous work on NGC\,6791 presented by Sanjayan \etl(2022a). Recently, Colman \etl(2022) presented the light curves of KIC targets observed in the \kep\ superstamps of NGC\,6819 using an image subtraction technique. We stress that the method and analysis presented in this work is independent of the one presented in Colman \etl(2022).
%, for instance our detrending was not as strong and allowed us to preserve some of the variations that are not present in the light curves derived by the authors.

%roadmap
In Section 2, we present a short description of the \kep\ mission and data processing in order to obtain the light curves of variable stars. In Section 3, we present our analysis of spectra taken at Nordic Optical Telescope (NOT), Apache Point Observatory (APO) and other spectra we found in public databases i.e. LAMOST survey, Apache Point Observatory Galactic Evolution Experiment (APOGEE) survey, MMT Hectospec 300 optical fiber fed spectrograph survey, and Apache Point Observatory 3.5-meter telescope survey. In Section 4, we invoke the \gaia\ astrometric parameters of variable stars we detected to establish their cluster membership. In section 5, we report the zoo of variable stars we found in the area of NGC\,6819. In Section 6, we present the results of the MIST\,(MESA Isochrones and Stellar Tracks) isochrones fitting to estimate the age of and the distance to NGC\,6819.

\section{Kepler Photometry}
During the original \kep\ mission four open clusters (NGC\,6791, NGC\,6819, NGC\,6811, NGC\,6866) were continuously observed. NGC\,6791 and NGC\,6819 were observed in the so-called "superstamps", which were used for observing multiple targets in crowded fields. The central most parts NGC\,6791 and NGC\,6819 were covered during Quarters (Q) 1--17. In the first part of our work, we reported a detailed study of the \kep\ superstamps of NGC\,6791 (Sanjayan \etl 2022a).

We downloaded the \kep\ superstamp data of NGC\,6819 from the Mikulski Archive for Space Telescopes\,(MAST\footnote{{https://archive.stsci.edu/}}). The data are 20\,x\,100 pixel boxes piled up in two contiguous 10 box stacks. The field of view of all pixels is 800\,x\,800\,arcseconds and covers the central most part of the cluster. The superstamps data are collected in the long cadence mode, lasting 30\,min. The pixel scale is 4\,arcsec. The data were collected over 1460\,days and are split into 18 quarters.

We searched for a flux variation by extracting fluxes for all time stamps in individual pixels for each of the quarters Q2\,--\,5. Next, a Fourier transform of the time-series data was performed in each pixel and each quarter separately. The pixels showing peaks (representing signal) in the amplitude spectra were selected. Signals that were identified with artifacts, either reported by Baran\,(2013) or those found in this project, were discarded. The optimal apertures are formed by contiguous pixels showing the same flux variation of S/N\,$\geq$\,5, not including pixels that contain too much signal from neighbors, even though the S/N ratio is still $\geq$\,5. Examples of optimal apertures can be seen in Figure\,6 of Sanjayan \etl(2022b). To keep the solar cells facing sunlight, every quarter the spacecraft rolled 90\,degrees, hence, with each quarterly rotation of the spacecraft, our targets landed on different CCD silicons. One CCD silicon covering the area of NGC\,6819 had failed prior to Q6 data collection, which caused quarterly data gaps during Q6,10 and 14. Using the optimal apertures for all targets showing flux variation, we used PyKE software (Kinemuchi \etl2012) to pull out the fluxes and correct them for instrumental artifacts by means of Co-trending Basis Vectors. Finally, using our custom scripts, we clipped the data at 4.5\,$\sigma$, detrended using spline fits, and normalized them to {\it parts per thousand} (ppt). The variable stars detected in our work is presented in Section\,5.

\section{Spectroscopy}
Using the Astrophysical Research Consortium's 3.5\,m telescope at APO, we obtained spectra for 5 stars. The spectra were collected from the Dual Imaging Spectrograph (DIS), using the B1200 and R1200 grating. In the blue channel, the grating was centered at 4\,300\,\AA\ and had a wavelength range of 1\,240\,\AA, while for the red channel, the grating was centered at 6\,400\,\AA\ and had a range of 1\,160\,\AA. The linear dispersions are 0.62\,\AA/pix in the blue and 0.58\,\AA/pix in the red. The magnitude range of the stars were 14.08 to 18.70 in the Johnson V, and for one object, the \gaia\ G magnitude was 17.39. The spectra were reduced and extracted using standard IRAF single slit spectroscopic techniques\,(Tody 1986, 1993). We observed the spectra of eight pulsating variable stars with the 2.56\,m NOT using ALFOSC spectrograph. In these spectra, we measured a dispersion of 2.2\AA\ from the full width at the half maximum of the lines in the arc spectra and a signal-to-noise ratio of $\approx$100 in the range of wavelengths 4\,950\,--\,5\,300\,\AA. For a number of targets we found spectra in the publicly available archives (links can be found in Table\,1). HECTOSPEC (Fabricant \etl 2005) operates at a resolution of approximately R$\approx$1\,000\,--\,2\,000, LAMOST (Zhao \etl 2012) offers low-resolution spectra with a resolution of around R$\approx$1\,800, while APOGEE (Ahn \etl 2014, Majewski \etl 2017) covers the H-band (1.51\,--\,1.70\,$\upmu$m) with a resolution of approximately R$\approx$22\,500.

We followed the fitting procedures from N\'emeth \etl(2012) in using {\sc XTgrid}. The $\chi$-square minimizing steepest-descent procedure fits an observation with interpolated {\sc Atlas}\,9 models, calculated in local thermodynamic equilibrium, drawn from the spectral library of Bohlin \etl(2017). Atmospheric parameters were derived by iteratively applying successive approximations of synthetic spectra to the observations. We interpolated in effective temperature (\teff), surface gravity (\logg) and scaled solar metallicity ([M/H]) space, and also fitted the radial velocity with respect to synthetic spectra. Iterations were pursued until the variation of surface parameters and the $\chi$-square decreased below 0.5\%. Uncertainties of parameters were determined by changing [M/H] and radial velocities in one dimension until the statistical limit for 60\% confidence was reached. The correlations between \teff\ and \logg\ were considered by the procedure. To avoid local minima the fitting procedure returns to the global minimization if a better solution is found during these calculations. Further details of the analysis methods are presented in Section 3 of Sanjayan \etl(2022a). Fig.\,1 shows the best-fit BOSZ/{\sc XTgrid} model to the NOT/ALFOSC observation of KIC\,5113357.

\begin{figure}
\includegraphics[width=\hsize]{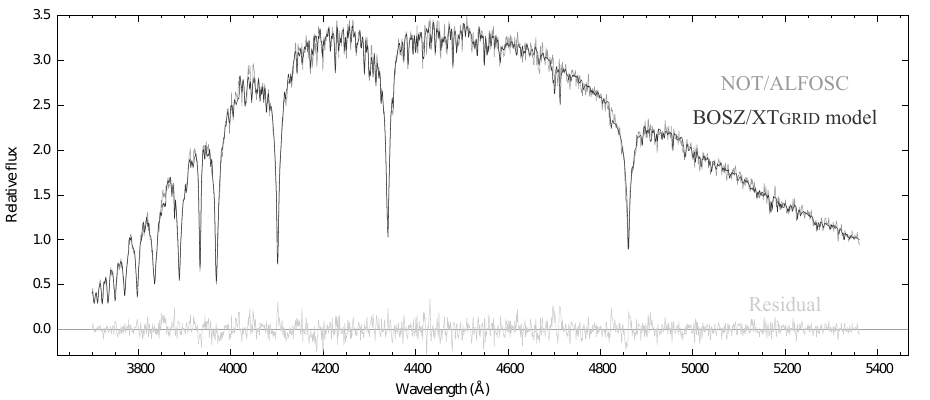}
\caption{Best-fit BOSZ/{\sc XTgrid} model to the NOT/ALFOSC observation of KIC\,5113357.}
\label{boszfit}
\end{figure}

\section{Cluster membership}
Our goal was to use the \gaia\ Data Release\,3 (DR3) astrometry to estimate the membership probability of all variable stars we found in this work. We downloaded astrometric data of over 36\,000 stars from the \gaia\ DR3 catalogue (\gaia\ Collaboration \etl 2022). We selected the stars in the area of NGC\,6819, defined by the cluster center $\alpha_{2000}$\,=\,19:41:17.2 and $\delta_{2000}$\,=\,+40:11:18 (Kamann \etl2019) and a tidal radius of 23 arcmin\,(Kalirai \etl2001). The membership analysis was done using five astrometric parameters, position ($\alpha$ and $\delta$ in equatorial coordinates), proper-motion ($\upmu_{\alpha}$ and $\upmu_{\delta}$) and, parallax $\pi$. To establish the membership of our variable stars we followed the method used and described in Sanjayan \etl(2022a). It employs variational Bayesian inference model with Dirichlet process prior (Ferguson 1973) using {\it scikit--learn} library in Python (Pedregosa \etl2011). We found 1\,971 \gaia\ targets to be cluster members, including 128 variable stars. A CMD of 15\,636 stars that were used in our analysis is presented in Fig.\,2. Majority of stars we found to be members of NGC\,6819 nicely mark the main sequence, red giant branch as well as BS and RC regions. 

We compared 1527 stars identified as the cluster members with a membership probability at least 0.7, reported by Cantat-Gaudin \etl(2018), with members found in our analysis. We found 1482 stars to be cluster members. Of the remaining 45 stars, we found 19 to be outside our search radius, we excluded two stars due to high uncertainty in the parallax and proper motion, while we found 24 stars not to be members.

RV data in the \gaia\ DR3 allow us to enhance the membership analysis for some of the stars in our sample, however the data can be contaminated by binarity, rotation and pulsation effects. The data need to be free of these effects before applying to the membership analysis to obtain a reliable solution. We preliminarily assumed the \gaia\ DR3 RV data are clean of those effects and we used an RV as a sixth parameter in our analysis. We recalculated the membership probability of 821 stars for which we collected the RV data, deriving 126 cluster members. The membership of these 126 stars have not changed after using the RV data. This outcome validates either our assumption that RVs have no additional effects or any such effects do not affect the membership analysis. Using the RV data of these 126 cluster members we estimated the average radial velocity of the cluster to be +3.5(5)\,km/s.

\begin{figure}
\includegraphics[width=\hsize]{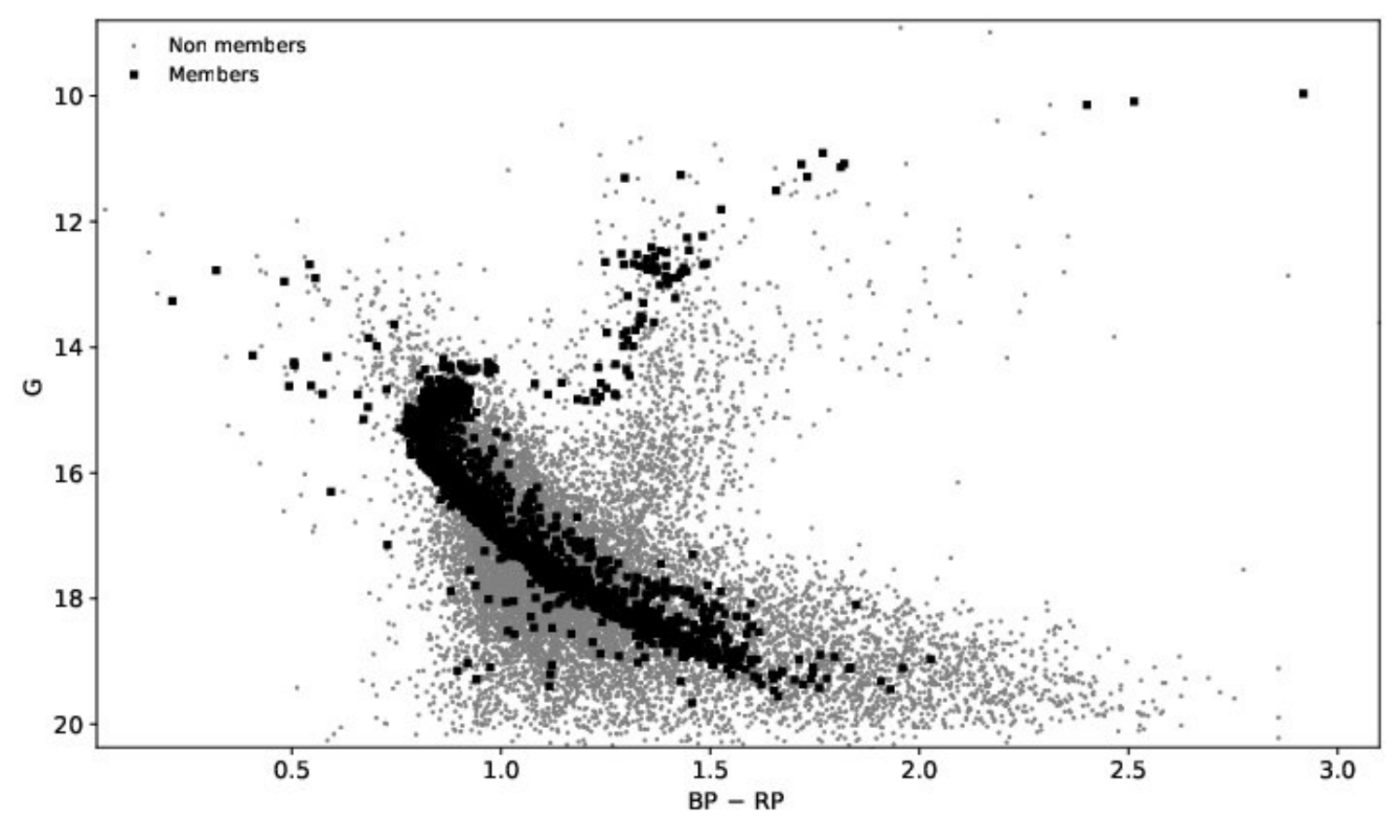}
\caption{Color-magnitude diagram of all stars included in our analysis.}
\label{membership}
\end{figure}

\section{A zoo of variable stars}
In total, we found 385 variable stars in the area of NGC\,6819. The observed variability is classified based on the light curve shape and the periods estimated from an amplitude spectrum. We distinguished three main variability types, {\it i.e}, binaries, pulsators and rotational. In addition, we listed several cases of unclassified and unidentified sources. Below we discuss each group of the flux variation. The cluster members and non members are listed in separate tables. For each type of variability we provide a number of newly detected variables.
%, not accounting for a report by Colman \etl(2022), since the authors did not discuss the variability survey, but only provided the light curves.

We used amplitude spectra calculated from the light curves to classify pulsators into solar-like, $\delta$\,Scuti, $\gamma$\,Dor, and semi-regular variables. Stars with a periodic flux variation but showing additional flux modulation are classified as rotational stars. A sample of the light curves with corresponding amplitude spectra for each type of variability and specific classes we detected in this analysis is presented in Fig.\,3.

We stress that our variability classification is subject to a number of caveats, which we have discussed in Sanjayan \etl(2022a) presenting similar work on NGC\,6791. Therefore the location of targets, especially binaries, in the CMDs and our RV estimates should be considered with caution.
 
\begin{figure}
\includegraphics[width=\hsize]{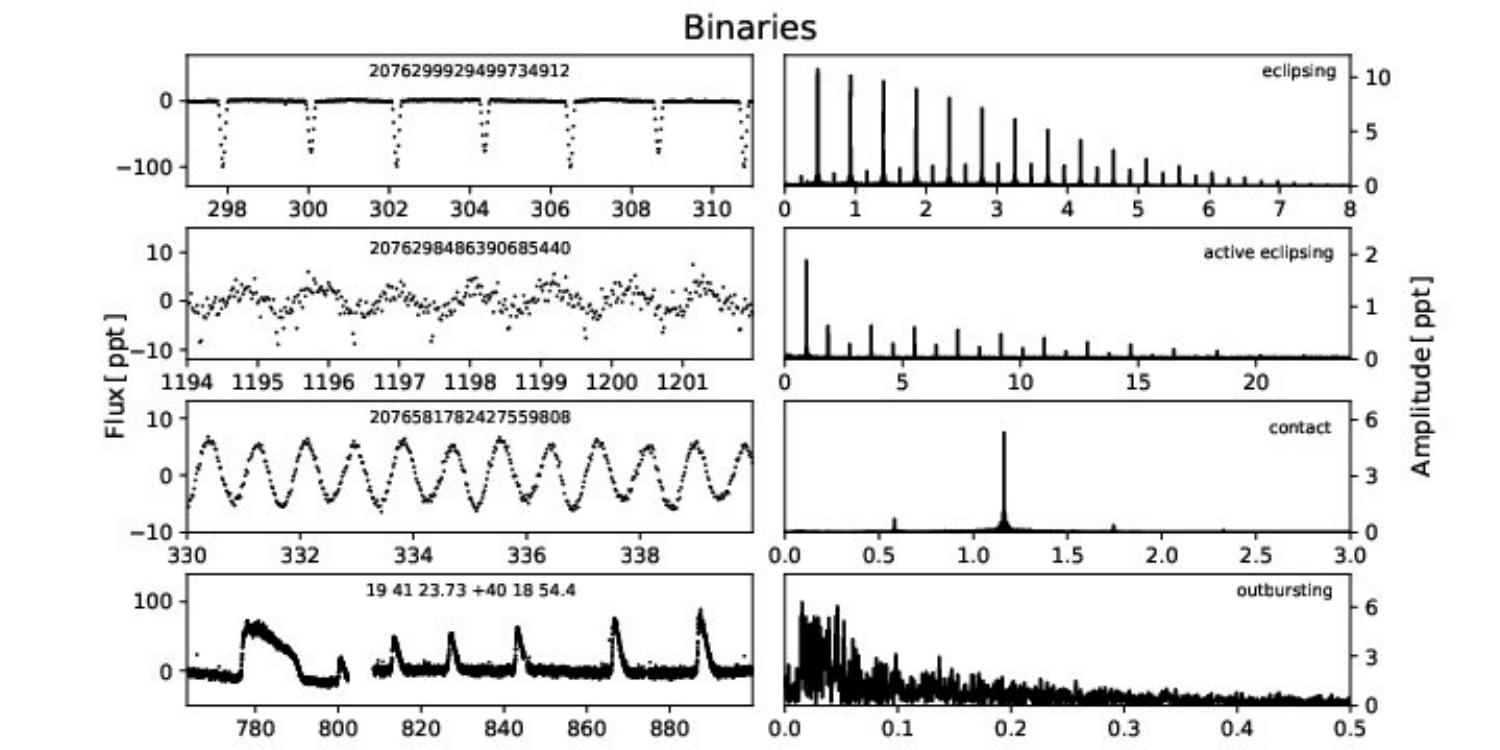}
\includegraphics[width=\hsize]{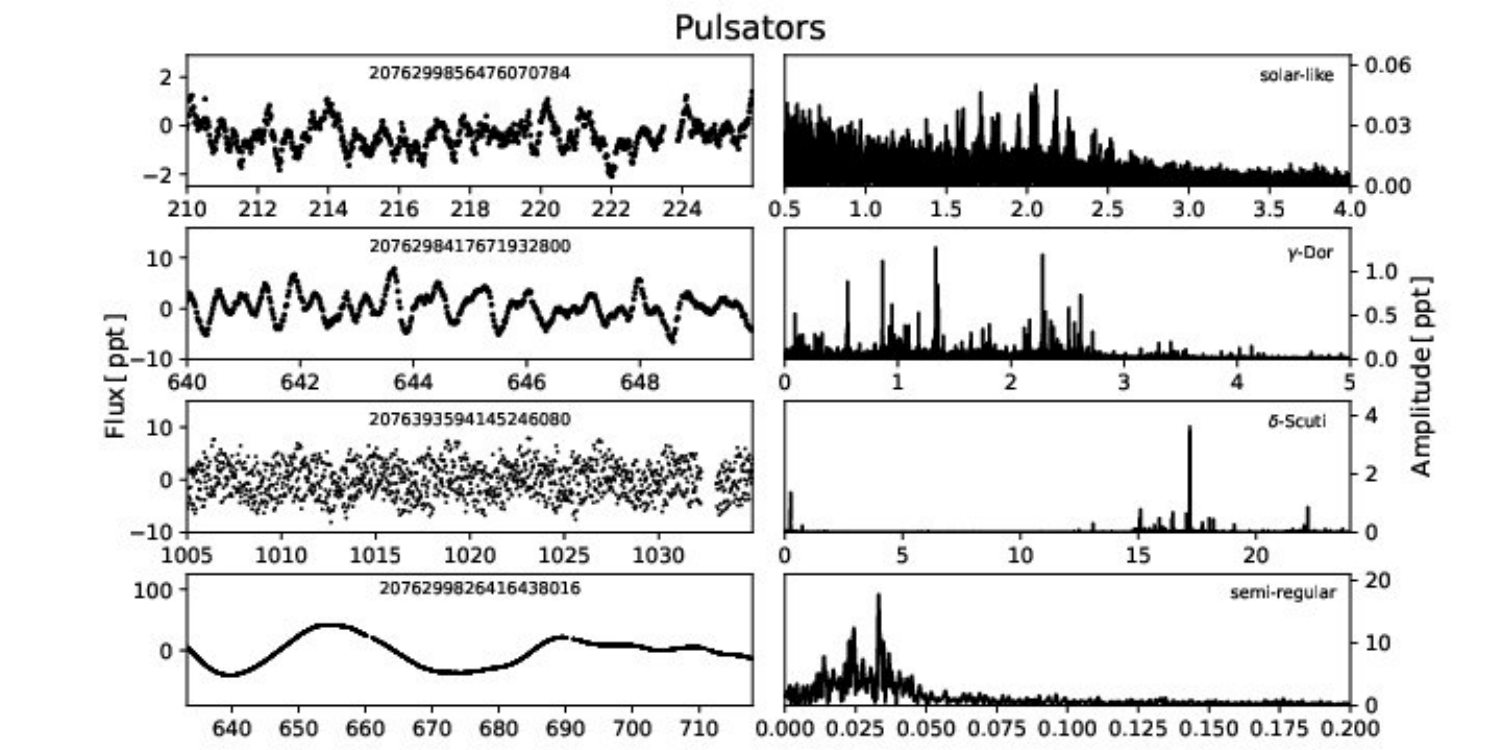}
\includegraphics[width=\hsize]{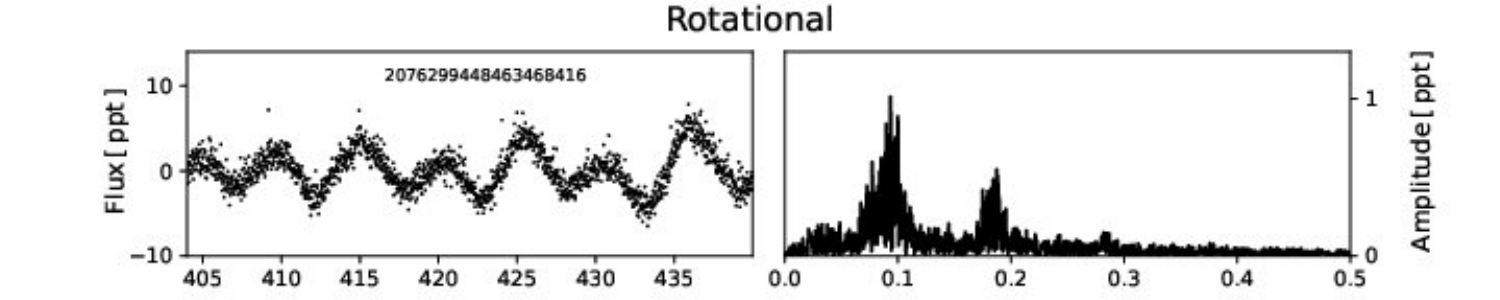}
\includegraphics[width=\hsize]{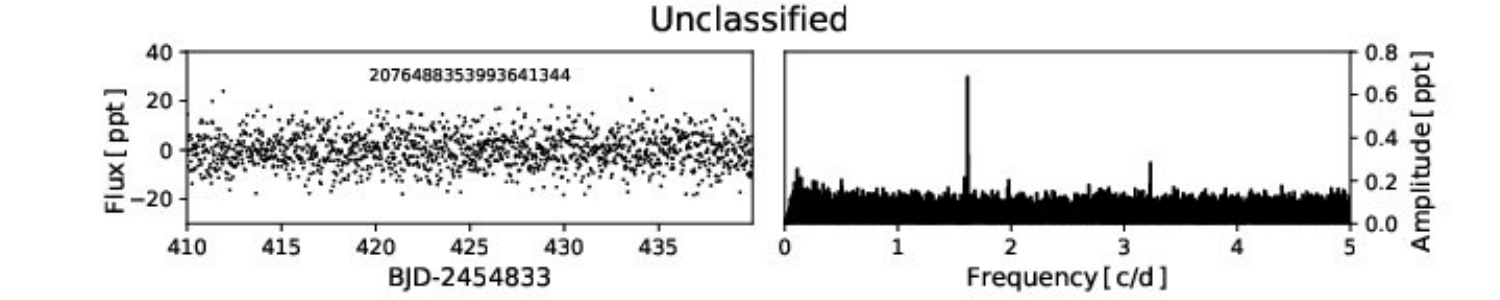}
\caption{A sample of the light curves and amplitude spectra for different types and classes of variable stars in the area of NGC\,6819.}
\label{lightcurves}
\end{figure}

\subsection{Binary variables}
We detected 59 binaries which included members, non members, and eight stars that are not associated with any \gaia\ targets, and hence their membership was not determined. From the phased light curves, we classified binaries into four classes, eclipsing, active eclipsing, contact and outbursting. We found 15 eclipsing systems with sharp eclipses and no out-of-eclipse variation. There are two members, 12 non-members, and one with no membership established. Eclipsing binaries with additional out-of-eclipse variation, likely caused by chromospheric activity, are called active eclipsing and we found eight members, 10 non-members and one with no membership established. The contact binaries are identified with systems showing light curves characteristics of W\,UMa system and we found seven members, 10 non-members and three with no membership established. Typical orbital periods of contact binaries are known not to be longer than 1 day so three stars may not be properly classified. Our classification is phenomenological, hence additional spectroscopic RV data may re-classify these objects. We found five outbursting stars, which we assumed to be binaries. Two of them are non-members while three have no membership established yet. In total we found 30 new variable binaries, including six cluster members, 17 field counterparts and seven with membership undetermined yet. We list the cluster members in Table\,1, non-members in Table\,5, while stars with no membership determined in Table\,10. In Fig.\,4 we present only phased light curves of the NGC\,6819 member binaries that we found to be new discoveries. In a CMD shown in the top panel of Fig.\,5, the majority of the binaries are located on the MS, while six are in the MSTO region. The latter ones can provide an independent estimate of the cluster age, if parameters are derived for individual components. As it was explained in Sanjayan \etl(2022a) the positions of binaries in the CMD should be treated with caution.

We identified five outbursting stars, which may possibly be associated with eruptions caused by a mass transfer. None of the outbursting stars were found to be members of the cluster. \gaia\ DR3\,2076392906942365568 was reported as an X\,ray cataclysmic variable during the XMM\,Newton survey by Gosnell \etl(2012).

Using the eclipses in eclipsing systems and the light maxima in contact binaries, we estimated the midtimes by means of the method described in Kwee \& van Woerden\,(1956). We used the midtimes to derive ephemerides, which are reported in Table\,1 and 5. For a few cases we provided only rough estimates of periods, which is caused by a low precision of data. In two cases we detected single eclipses only. To search for an orbital period variation we calculated Observed minus Calculated (O-C) diagrams. We found significant orbital period variations in five binaries, including only one cluster member (Fig.\,6). In two cases, KIC\,5112759 and KIC\,5113461, the variations may look sinusoidal caused by an additional body in these systems. The other three cases are not easy to interpret.

%Binary - table
%\begin{center}
%\input{table1.tex}
%\end{center}
%%%%%%%%%%%%%%%%%%%%%%%%%%%%%%%%%%%%%%%
%Table of BINARIES IN CLUSTER
\begin{sidewaystable}
\caption{List of binary stars that are cluster members. The newly discovered variables are marked in bold in the \gaia\ DR3 column. The data availability from MAST is denoted in the superscript of the KIC\,ID, where X means no data, L and S stands for the LC and SC data, respesctively; the numbers before the letters represent how many quarters of data are available. CMD refers to the position in the color-magnitude diagram (MS - main sequence, MSTO - main sequence turn off, RGB - red giant branch, BS - blue straggler, AGB - asymptotic giant branch, RC- red clump, HB - horizontal branch, EHB - extreme horizontal branch). The superscript $^e$ of the period denotes significantly eccentric orbit. The superscript $^+$ of the CMD indicates that not all five astrometric parameters were used in the membership analysis. HRD refers to a position in the T$_{\rm eff}$, $\log g$ diagram. The source of the spectra is marked in the Ref column, i.e. 1 - HECTOSPEC, 2 - NOT, 3 - LAMOST, 4 - APO35, and 5 - APOGEE. The $^*$ indicates that the atmospheric parameters are derived in this work.}
\label{tab:cluster_binaries}
\centering
\resizebox{\columnwidth}{!}{
\begin{tabular}{clllccclllll}
\hline\hline
\multicolumn{1}{c}{\multirow{2}{*}{\gaia\ DR3}} & \multicolumn{1}{c}{\multirow{2}{*}{KIC}} & \multicolumn{1}{c}{P$_{\rm orb}$} & \multicolumn{1}{c}{T$_0$} & \multicolumn{1}{c}{G} & \multirow{2}{*}{CMD} &  \multirow{2}{*}{HRD} & \multicolumn{1}{c}{T$_{\rm eff}$} & \multicolumn{1}{c}{\multirow{2}{*}{\logg}} & \multicolumn{1}{c}{RV} & \multicolumn{1}{c}{\multirow{2}{*}{[Fe/H]}} & \multicolumn{1}{c}{\multirow{2}{*}{Ref}}\\
&& \multicolumn{1}{c}{[days]} & \multicolumn{1}{c}{[BJD]} & \multicolumn{1}{c}{[mag]} &&& \multicolumn{1}{c}{[K]} && \multicolumn{1}{c}{[km/s]} &&\\
\hline\hline
\multicolumn{12}{c}{eclipsing}\\
{\bf 2076300062631198208} & 5024475\,$^{X}$	     & 0.35461351(48) & 2\,454\,964.3935(11)  & 16.950	& MS & -- &	--	&  --  &  --  & -- & --  \\
2076298864347818624 & 5024447\,$^{14L}$    & 771.8122(14)	  & 2\,455\,267.6048(10)  & 14.933	& MS  & -- &	--	& --  & --  & --  & --  \\	
\hline
\multicolumn{12}{c}{active eclipsing}\\
2076300101298442368 & 5112456\,$^{X}$	     & 1.04163866(35) & 2\,454\,965.06811(26) & 16.943	& MS  & MS & 5\,660(200) &	4.36(1) & 0(13) & -0.439(31) & 4\,$^*$   \\
2076299929499734912 & 5024292\,$^{1S, 4L}$ & 4.3010306(16)  & 2\,454\,967.449238(30) & 15.001	& MS  & -- & -- & -- & -- & -- & -- \\
2076299826420588288 & 5024450\,$^{1S, 11L}$& 3.0518491(12)  & 2\,455\,185.42743(27) & 14.985	& MS  & -- &	--	& --  & --  & --  & --  \\
2076487121349914752 & 5023948\,$^{1S, 9L}$ & 3.64930727(38) & 2\,455\,465.83818(6)  & 14.991	& MS  & -- & -- & -- & -- & -- & --  \\	
2076299169277916160 & 5024064\,$^{X}$      & 2.05541020(25) & 2\,454\,965.97867(10) & 18.528	& MS	& -- &	--	& --  & --  & --  & --  \\
{\bf 2076298486390685440} & 5024364\,$^{X}$	     & 1.089253(5)	  & 2\,454\,965.8187(39)  & 16.982	& MS & -- &	--	& --  & --  & --  & --  \\
2076393422338668800 & 5113176\,$^{X}$      & 2.50489870(38) & 2\,454\,964.84980(12) & 18.973	& MS	& -- &	--	& --  & --  & --  & --  \\
{\bf 2076392838230907264} & 5024980\,$^{X}$      & 414.54921(33)$^e$  & 2\,455\,081.23049(61) & 17.427	& MS & -- &	--	& --  & --  & --  & --  \\
\hline
\multicolumn{12}{c}{contact}\\
2076581713708050176 & 5112407\,$^{X}$	     & 0.366025967(25)& 2\,454\,964.90170(5)  & 16.415	& MS	& MS &	5\,620(50) & 4.345(37) & 65(3) & -1.15(8) & 1\,$^*$           \\
2076299792060826496 & 5024624\,$^{X}$	     & 0.30321101(8)  & 2\,454\,964.70244(22) & 17.097	& MS	& MS &	5\,590(20) & 4.865(23) & 26(2) & -0.430(23)  &   1\,$^*$      \\
2076299448463338368 & 5023989\,$^{X}$      & 0.26367746(8)  & 2\,454\,964.78122(23) & 19.508	& --$^+$	     & -- &	--	& --  & --  & --  & --  \\
{\bf 2076581782427559808} & 5112708\,$^{X}$      & 1.7195289(19)  & 2\,454\,965.2048(9)	  & 14.915	& MS & -- &	--	& --  & --  & --  & --  \\	
2076394105234042496 & 5112759\,$^{X}$      & 0.25622643(8)  & 2\,454\,964.66578(25) & 17.788	& MS	& MS &	4\,650(30) &	4.184(28) &	15(5)	& -0.67(7) &   1\,$^*$     \\
{\bf 2076389608415499392} & 5025105\,$^{X}$      & 1.5747014(35)  & 2\,454\,965.1646(19)  & 15.286	& MS & -- &	--	& --  & --  & --  & --  \\	
{\bf 2076300066938740096} & 5112693\,$^{X}$	     & 0.211752(1)	  & 2\,454\,964.59610(38) & 16.828	& MS & -- &	--	& --  & --  & --  & --  \\
%131-2076300131353981312 & 5112536\,$^{X}$      & 0.6732      &                     &17.686   & MS &  & 4600(40)	& 4.33(34) & 220(1)	    & 0.718(0.032) &   \\
\hline\hline
\end{tabular}
}
\flushleft
\footnotesize{HECTOSPEC - https://oirsa.cfa.harvard.edu, APOGEE - https://www.sdss.org/dr16/irspec, LAMOST - http://dr6.lamost.org, APO\,35 - https://www.apo.nmsu.edu}
\end{sidewaystable}
%%%%%%%%%%%%%%%%%%%%%%%%%%%%%%%%%%%%%%%%%%%%%%%%%

% phased LC cluster member binaries
\begin{figure}
\includegraphics[width=\hsize]{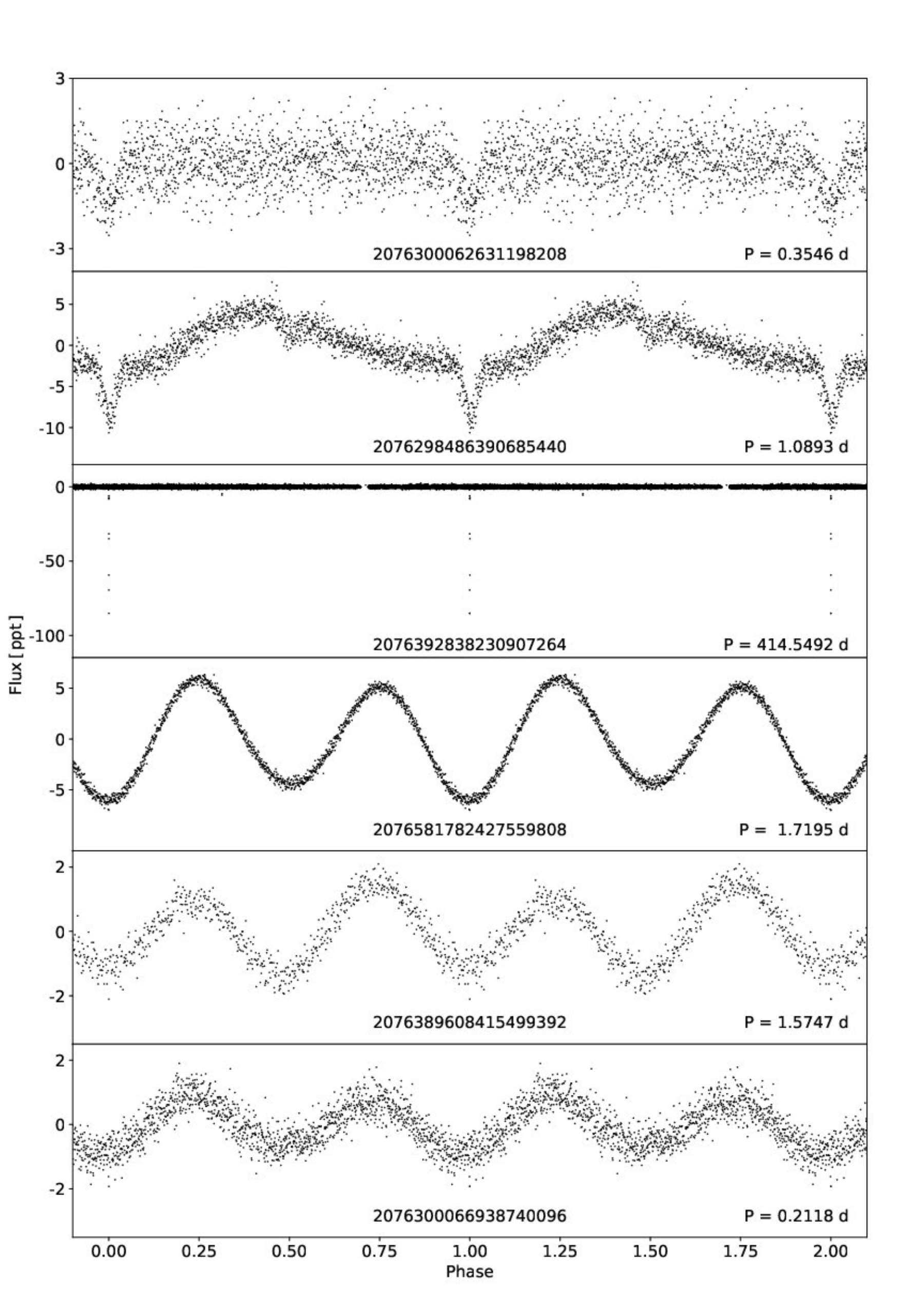}
\caption{Phased light curves of newly found cluster member binaries. See Table\,1 for details.}
\label{phasedlightcurves}
\end{figure}

\begin{figure}
\centering
\includegraphics[width=0.8\hsize]{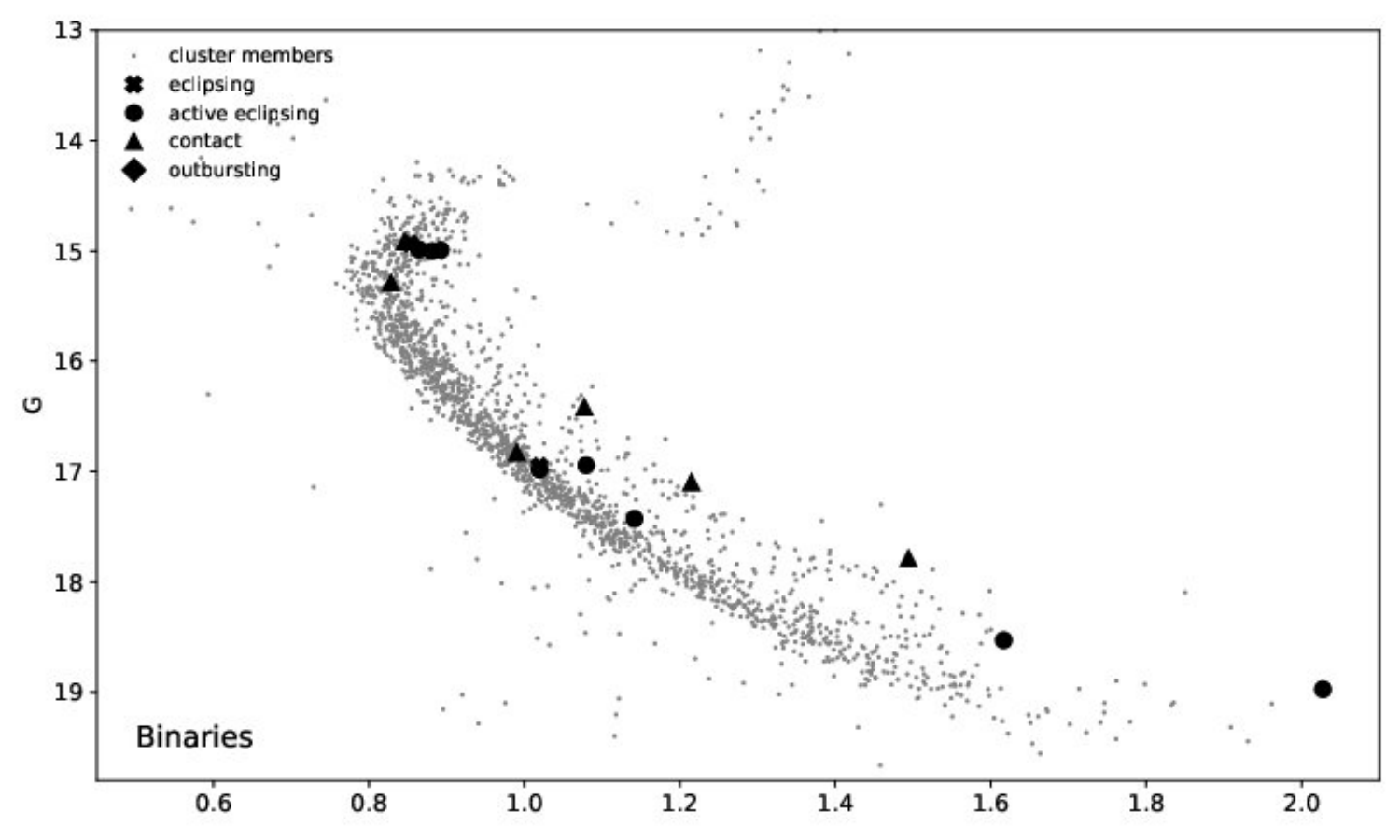} \\
\includegraphics[width=0.8\hsize]{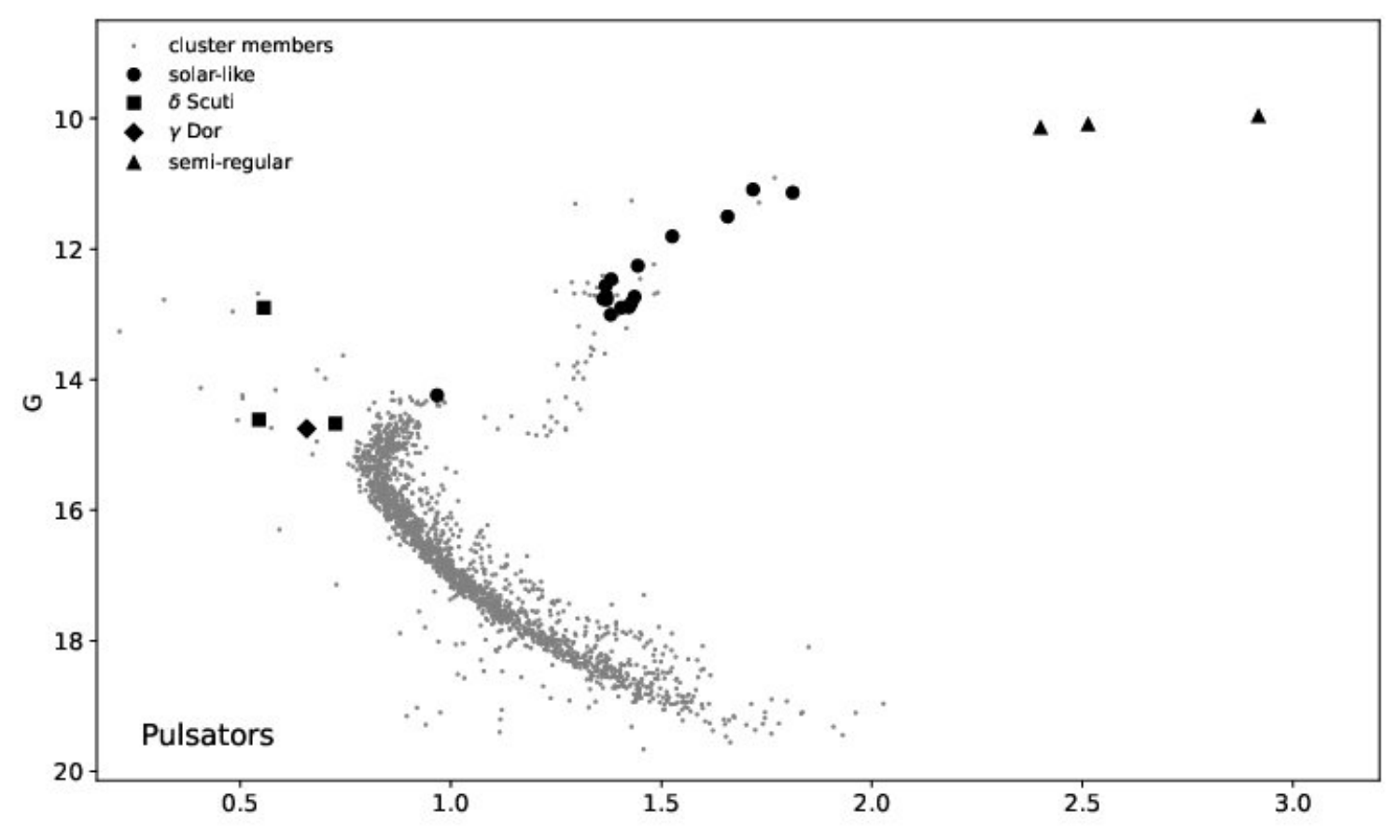} \\
\includegraphics[width=0.8\hsize]{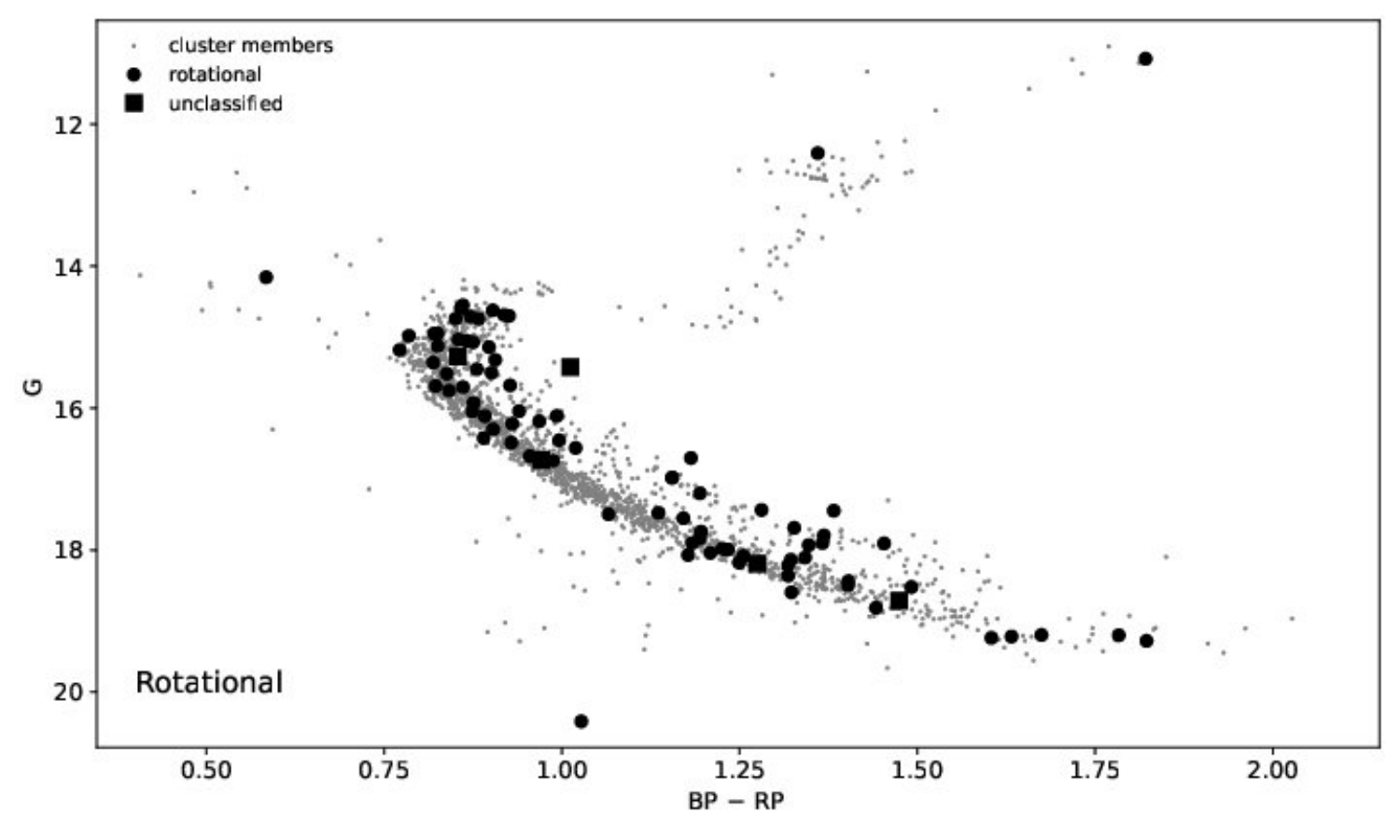}
\caption{Color-magnitude diagram of NGC\,6819. Binaries, pulsators, and rotational/unclassified cluster members are marked in the top, middle, and bottom panels, respectively.}
\label{cmd}
\end{figure}

% oc of binaries
\begin{figure*}
\includegraphics[width=\hsize]{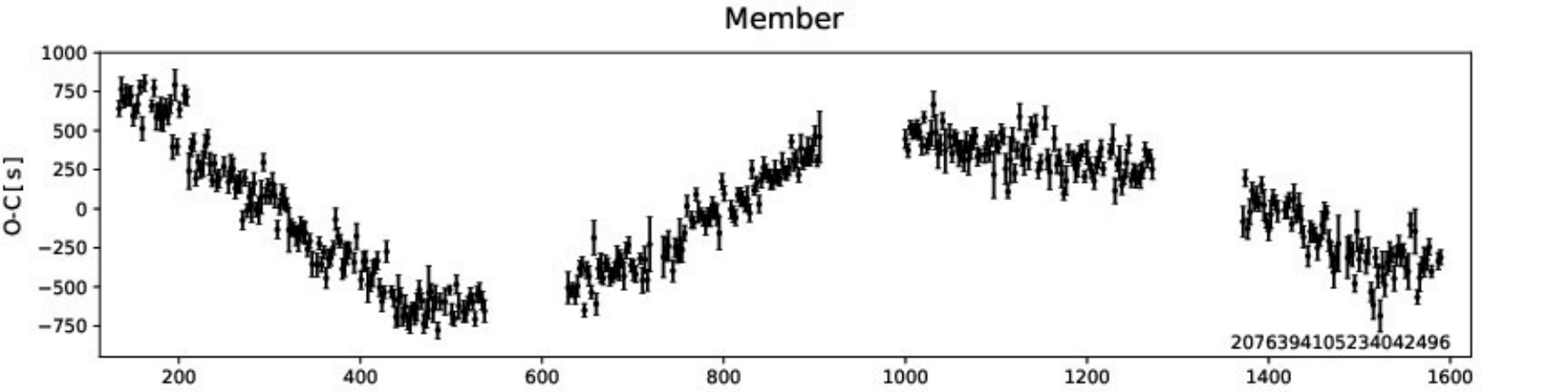}
\includegraphics[width=\hsize]{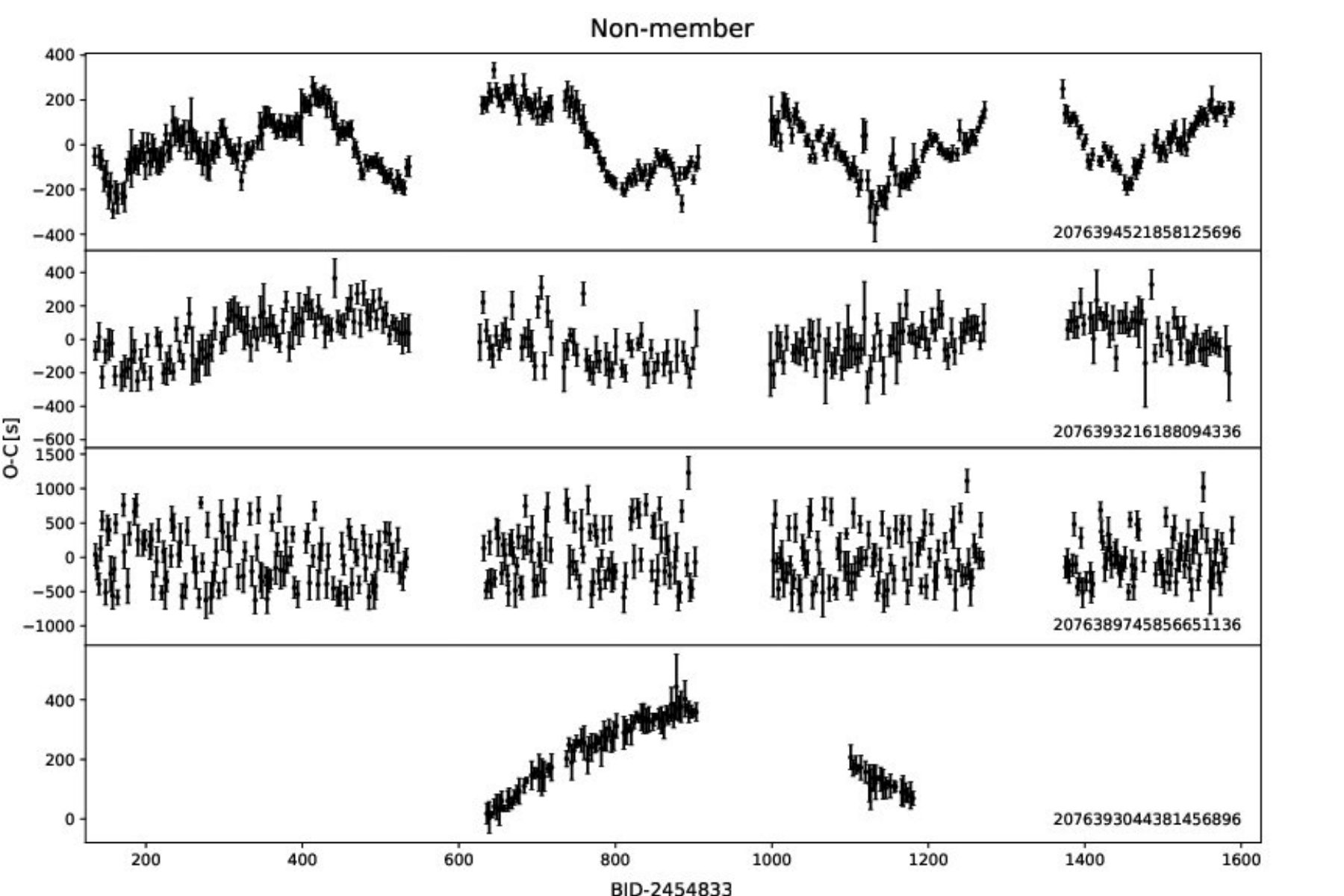}
\caption{O-C diagrams of four contact and one active eclipsing systems showing significant orbital period variations. \gaia\ DR3 numbers are given in the bottom right corners.}
\label{omc}
\end{figure*}

\subsection{Pulsators}
We found a flux variation in 69 stars, which we interpreted as stellar oscillations and we separated the following classes, solar-like, $\delta$\,Scuti, $\gamma$\,Doradus and semi-regular pulsators. We found 24 pulsating cluster members including 17 solar-like, three $\delta$\,Scuti, one $\gamma$\,Doradus and three semi-regular. The field counterpart includes, 34 solar-like, two $\delta$\,Scuti, six $\gamma$\,Doradus and three semi-regular pulsators. In total we found 32 new variable pulsators, including two cluster members and 30 field counterparts. We present the pulsating cluster members in Table\,2 and the field counterpart in Table\,6. The location of the former group in the CMD is shown in the middle panel of Fig.\,5. A detailed analysis of solar-like and $\delta$\,Scuti/$\gamma$\,Doradus pulsators will be reported by Themessl \etl(in prep.) and Guzik \etl(2023), respectively.

The majority of pulsating cluster members are located on the RGB, six are the RC objects, while four $\delta$\,Scuti and $\gamma$\,Doradus objects seems to be located in the BS region. The semi-regular pulsators are the asymptotic giant branch (AGB) objects.

Since this work is also a part of our search for pulsating extreme horizontal branch stars, we specifically looked for any flux variation that is characteristic of pulsating hot subdwarfs identified in NGC\,6791 and reported by Sanjayan \etl(2022a) and (2022b). Hot subdwarfs can be formed through either a degenerate or a non-degenerate channel. The former channel works only for progenitor masses up to around 2.2\,solar masses. Such stars need significantly less time than 2.5\,Gyr to settle down onto the HB and even to reach the end of the white dwarf (WD) cooling tracks, becoming too faint to be detected. The latter channel is available for progenitors more massive than around 2.2\,solar masses. Given the age of the cluster, stars with masses in a range of 1.5\,--\,1.6\,solar masses had enough time to evolve to the HB and, if most of the hydrogen is lost, to become hot subdwarfs. More massive stars could have evolved to become post-HB stars (the most massive ones can now be very cool WDs and too faint to be seen), while less massive stars are still on the RGB or the MS, if less massive than 1.45\,solar masses. Our consideration leads to a conclusion that the CMD of NGC\,6819 should be well populated with stars all the way from the MS to the WD stage. This is not what we can see in Fig.\,5. There is neither the WD nor the extreme HB populations. We stress that our consideration is based on a single star evolutionary time scale. Binary channels contributing either to a mass loss or a merger event can extend the time scale significantly, which could explain a lack of the hot subdwarf population in this cluster.
 
%\begin{center}
%\input{table2.tex}
%\end{center}
%%%%%%%%%%%%%%%%%%%%%%%%%%%%%%%%%%%%%%%%%%%%%%%%%%%%%
%Table of PULSATORS IN CLUSTER
\begin{table*}
\caption{List of cluster members showing pulsations. See caption of Table\,1 for explanation.}
\label{tab:cluster_binaries}
\centering
\renewcommand{\arraystretch}{1.2}
\resizebox{\columnwidth}{!}{%
\begin{tabular}{clccclllll}
\hline\hline
\multicolumn{1}{c}{\multirow{2}{*}{\gaia\ DR3}} & \multicolumn{1}{c}{\multirow{2}{*}{KIC}} & \multicolumn{1}{c}{G} & \multirow{2}{*}{CMD}& \multirow{2}{*}{HRD} & \multicolumn{1}{c}{T$_{\rm eff}$} & \multicolumn{1}{c}{\multirow{2}{*}{\logg}} & \multicolumn{1}{c}{RV} & \multicolumn{1}{c}{\multirow{2}{*}{[Fe/H]}} & \multicolumn{1}{c}{\multirow{2}{*}{Ref}}\\
&& \multicolumn{1}{c}{[mag]} &&& \multicolumn{1}{c}{[K]} && \multicolumn{1}{c}{[km/s]} &&\\
\hline\hline
\multicolumn{10}{c}{solar-like}\\
2076488427020005632 &	5111940\,$^{14L}$     & 13.005	& RGB & RGB	& 4\,771(92)	& 2.79(11)	& 2.964(71)	& 0.200(28) & 5  \\
2076581919866484224 &	5112401\,$^{15L}$     & 12.562	& RC & RC	& 4\,842(92)	& 2.65(11)	& 2.488(60)	& 0.200(27) & 5  \\
2076487872957025408 &	5112387\,$^{15L}$     & 12.776	& RC& RC & 4\,839(92)	& 2.85(11)	& 2.80(22)	& 0.200(28)	& 5  \\
2076581713708596352 &	5112481\,$^{15L}$     & 14.239	& RGB & RGB	& 4\,152(69)	& 1.66(8)	&-1.664(80)	& 0.000(25) & 5  \\
2076487838597288320 &	5112373\,$^{15L}$     & 12.755	& RC & RC	& 4\,821(92)	& 2.73(11)	& 2.29(51)	& 0.200(28)	& 5  \\
2076300101298433792 &	5024297\,$^{14L}$     & 12.896	& RGB &--	& --			& --          & --          & --          & --  \\
2076299792060819328 &	5024583\,$^{14L}$     & 12.732	& RGB & RGB	& 4\,687(92)	& 2.55(11)	& 1.9843(34)& 0.200(27) & 5	 \\
{\bf 2076299856476070784} &	5024511\,$^{X}$       & 12.256	& RGB &--	& 	--		&   --        &     --      &      --     &--    \\
%{\bf 2076299414103588224} &	5023913\,$^{X}$       & 15.274	& MS &--	& 	--		&    --       &        --   &    --       & --   \\ %moved to unclassified
2076394659297111552 &	5113061\,$^{15L}$     & 11.088	& RGB &	RGB & 4\,221(92)	& 1.67(11)	& 2.56(11)	& 0.100(26) & 5  \\
2076581816787312896 &	5112744\,$^{15L}$     & 12.890	& RGB & RGB	& 4\,691(92)	& 2.63(11)	& 2.7996(89)& 0.100(28) & 5	 \\
2076393937742523904 &	5112730\,$^{15L}$     & 12.719	& RC & RC	& 4\,823(92)	& 2.72(11)	& 1.350(75) & 0.200(27) & 5  \\
2076393692914035456 &	5112786\,$^{15L}$     & 11.502	& RGB &	RGB & 4\,308(92)	& 1.86(11)	& 3.824(45) & 0.100(27) & 5  \\
2076393869023076608 &	5112948\,$^{14L}$     & 12.833	& RGB & RGB	& 4\,766(92)	& 2.68(11)	& 3.588(24) & 0.200(27) & 5  \\
2076393765943846656 &	5112938\,$^{14L}$     & 12.762	& RC &	RC & 4\,842(92)	& 2.69(11)	& 1.328(28) & 0.200(27) & 5  \\
2076393624194561152 &	5024750\,$^{15L}$     & 11.806	& RGB & RGB	& 4\,501(92)	& 2.11(11)	& 2.177(27) & 0.100(27) & 5  \\
2076299860780325504 &	5024601\,$^{14L}$     & 12.465	& RC & --	& 	--		&      --     &    --       &      --     &    --\\
2076298761268676352 &	5024851\,$^{15L}$     & 11.136	& RGB &	RGB & 4\,159(92)	& 1.60(11)	& 6.23(23)  & 0.100(26) & 5  \\

\hline
\multicolumn{10}{c}{$\delta$-Scuti}\\
2076299826420542080 &	5024468\,$^{X}$       & 12.898	& BS &	BS & 7\,770(50)	& 4.38(10)	& -14(2) &	-0.2(1) & 2\,$^*$    \\
2076299482823088000 &	5024084\,$^{1S, 14L}$ & 14.674	& BS & --	& 	--		&      --     &    --      &  --   & --  \\
{\bf 2076393594145246080} &	5113357$^{X}$     & 14.613	& BS & BS & 7\,270(50)	& 3.70(5)	& 65(3)   & -0.53(10)   & 2\,$^*$    \\
\hline
\multicolumn{10}{c}{$\gamma$-Dor}\\
2076298417671932800 &	5024455\,$^{1S, 14L}$ & 14.751	& BS & -- & 	--		& 	  --      &    --       &  --    & --     \\
\hline
\multicolumn{10}{c}{semi-regular}\\
2076299826416438016 &	5024470\,$^{X}$	      &  9.964	    & AGB &	-- & 	--		&    --      &   --   &  --   & --\\
2076582950658667264 &	5199859\,$^{15L}$ 	  &  10.143	    & AGB & AGB & 3\,828(92)	& 1.00(11) &  2.1875	& -0.200(27) & 5  \\
2076394728016615680 &	5113517\,$^{1L}$  	  &  10.091	& AGB &	-- &	--		&      --    &     --       &      --     & -- \\
\hline\hline
\end{tabular}
}
\end{table*}
%%%%%%%%%%%%%%%%%%%%%%%%%%%%%%%%%%%%%%%%%%%%%%%%%%%%%%%%%%%%%%%%%%

\subsection{Rotational Variables}
Stars showing modulated periodic flux variations are classified as rotational variables. Such a flux modulation can also be verified by amplitude spectra showing complex peak profile, as a consequence of amplitude/frequency variation. The rotational variability is usually identified with migrating star spots on the stellar surface. Our classification of rotational variables may not always be correct, since it can be mimicked {\it e.g.} by small amplitude ellipsoidal variability or unstable long-period pulsations. Light curves showing stable periodic flux variations are classified as binaries, since we do not expect non migrating star spots.

In total, we found 233 variable stars showing rotational variability, which makes them the most abundant variable population in NGC\,6819. There are 82 cluster members (Table\,3), 136 field counterparts (Tables\,7 and 8), and 15 stars with no membership established (Table\,10). The periods of rotational variables are in a wide range of 0.2\,--\,45\,days. Cross matching our list of rotational variables with the International Variable Star Index (VSX) catalogue, we found 15 rotational variables to be of BY\,Draconis type, while KIC\,5112508 is a rotating X\,ray binary reported by Gosnell \etl(2012). In summary, we found 185 new variable rotating stars, including 70 cluster members, 100 field counterparts and 15 with membership undetermined yet. The location of the cluster members in the CMD is shown in the bottom panel of Fig.\,5.

%\begin{center}
%\input{table3.tex}
%\end{center}
%%%%%%%%%%%%%%%%%%%%%%%%%%%%%%%%%%%%%%%%%%%%%%%%%%%%%
%Table of Rotational variables IN CLUSTER
\begin{table*}
\caption{List of cluster members showing rotational variability. See caption of Table\,1 for explanation.}
\label{tab:cluster_rotational}
\centering
\resizebox{\columnwidth}{320pt}{%
\begin{tabular}{clrccclllll}
\hline\hline
\multicolumn{1}{c}{\multirow{2}{*}{\gaia\ DR3}} & \multicolumn{1}{c}{\multirow{2}{*}{KIC}} & \multicolumn{1}{c}{Period} & \multicolumn{1}{c}{G} & \multirow{2}{*}{CMD} & \multirow{2}{*}{HRD} & \multicolumn{1}{c}{T$_{\rm eff}$} & \multicolumn{1}{c}{\multirow{2}{*}{\logg}} & \multicolumn{1}{c}{RV} & \multicolumn{1}{c}{\multirow{2}{*}{[Fe/H]}} & \multicolumn{1}{c}{\multirow{2}{*}{Ref}}\\
&& \multicolumn{1}{c}{[days]} & \multicolumn{1}{c}{[mag]} &&& \multicolumn{1}{c}{[K]} && \multicolumn{1}{c}{[km/s]} &&\\
\hline\hline
{\bf 2076489148574565888} &	5112059\,$^{X}$	      & 22.3010	&  17.901	& MS & -- & -- & -- & -- & -- & -- \\ 
{\bf 2076489217294036352} &	5199551\,$^{X}$	      & 23.1662	&  17.495	& MS & -- & -- & -- & -- & -- & -- \\ 
2076488942416130688 &	5112104\,$^{30S,11L}$ &	5.6490	&  14.593	& MSTO & -- & -- & -- & -- & -- & -- \\ 
{\bf 2076488869389772800 } & 5112151\,$^{X}$ & 18.5003 & 18.812 & MS & -- & -- & -- & -- & -- & -- \\
2076492034792541952 &	5111815\,$^{14S}$     & 3.5750	&  14.948	& MS & -- & -- & -- & -- & -- & -- \\ 
{\bf 2076488495739451136} &	5111800\,$^{X}$	      & 3.8679	&  15.181	& MS & -- & -- & -- & -- & -- & -- \\ 
{\bf 2076488770617529088} &	5112068\,$^{X}$	      & 15.9215	&  14.982	& MS & -- & -- & -- & -- & -- & -- \\ 
{\bf 2076488152142173312} &	5112298\,$^{X}$	      & 2.3639	&  15.688	& MS & -- & -- & -- & -- & -- & -- \\ 
{\bf 2076488319633886336} &	5111907\,$^{X}$	      & 15.7120	&  18.175	& MS & -- & -- & -- & -- & -- & -- \\ 
{\bf 2076488289581014400} &	5111849\,$^{1S, 2L}$  & 5.9651	&  14.551	& MSTO & -- & -- & -- & -- & -- & -- \\ 
{\bf 2076488147835261696} &	5112257\,$^{X}$	      & 19.5174	&  18.594	& MS & -- & -- & -- & -- & -- & -- \\ 
{\bf 2076488903749503104} &	5112187\,$^{X}$	      & 19.6927	&  19.217	& MS$^+$ & -- & -- & -- & -- & -- & -- \\ 
{\bf 2076487945983692672} &	5112173\,$^{X}$	      & 7.1660	&  16.702	& MS & -- & -- & -- & -- & -- & -- \\ 
{\bf 2076487568026522880} &	5111983\,$^{X}$	      & 6.1842	&  17.794	& MS & -- & -- & -- & -- & -- & -- \\ 
{\bf 2076300135658207232} &	5112512\,$^{X}$	      & 5.5446	&  16.422	& MS & -- & -- & -- & -- & -- & -- \\ 
{\bf 2076300131350662912} &	5112490\,$^{X}$	      & 7.5360	&  16.452	& MS & -- & -- & -- & -- & -- & -- \\ 
{\bf 2076393972102236800} &	5112566\,$^{X}$	      & 11.0660	&  17.832	& MS & -- & -- & -- & -- & -- & -- \\ 
{\bf 2076487808537778688} &		 --    	          & 0.9752	&  19.236	& MS & -- & -- & -- & -- & -- & -- \\ 
{\bf 2076300101298445184} &	5112431\,$^{X}$	      & 5.2943	&  16.739	& MS & -- & -- & -- & -- & -- & -- \\ 
{\bf 2076300066938738176} &	5112630\,$^{X}$	      & 4.4602	&  15.141	& MS & -- & -- & -- & -- & -- & -- \\ 
2076300131353981312 &	5112536\,$^{X}$	          & 0.6732	&  17.686	& MS & -- & -- & -- & -- & -- & \\ 
2076300101298455168 &	5112519\,$^{X}$	      & 5.9061	&  17.900	& MS & -- & -- & -- & -- & -- & -- \\ 
{\bf 2076300101299126912} &	5112434\,$^{X}$	      & 5.0307	&  14.740	& MSTO & -- & -- & -- & -- & -- & -- \\ 
{\bf 2076393903382760576} &	5112677\,$^{X}$	      & 7.1078	&  14.707	& MSTO & -- & -- & -- & -- & -- & -- \\ 
2076299826420536320 &	5024456\,$^{14L}$     & 3.2313	&  11.079	& RGB	& RGB & 4\,062(69)	& 1.49(8) & 1.383(76) & 0.000(25)& 5 \\
{\bf 2076487082683547008} &	5023857\,$^{X}$	      & 11.5019	&  18.360	& MS$^+$	 & -- & -- & -- & -- & -- & -- \\ 
{\bf 2076300066938722176} &	5024480\,$^{X}$	      & 4.1951	&  15.679	& MS	 & -- & -- & -- & -- & -- & -- \\ 
{\bf 2076299585902316416} &	5024114\,$^{X}$	      & 11.9194	&  15.754	& MS & -- & -- & -- & -- & -- & -- \\ 
{\bf 2076299723341297920} &	5024318\,$^{X}$	      & 3.7930	&  15.516	& MS & -- & -- & -- & -- & -- & -- \\ 
{\bf 2076299929499804288} &	5024287\,$^{1S, 1L}$  & 7.1699	&  14.623	& MS & -- & -- & -- & -- & -- & -- \\ 
{\bf 2076299963859447936} &	5024151\,$^{X}$	      & 45.9273	&  14.157	& BS & -- & -- & -- & -- & -- & -- \\ 
{\bf 2076299448463468416} &	5023967\,$^{X}$	      & 10.6964	&  16.676	& MS & -- & -- & -- & -- & -- & -- \\ 
{\bf 2076299757701068800} &	5024600\,$^{X}$	      & 3.9612	&  16.221	& MS & -- & -- & -- & -- & -- & -- \\ 
{\bf 2076299723341302272} &	5024403\,$^{X}$	      & 5.0020	&  15.704	& MS & -- & -- & -- & -- & -- & -- \\ 
{\bf 2076299036142588032} &	5024760\,$^{X}$	      & 2.4474	&  16.043	& MS & -- & -- & -- & -- & -- & -- \\ 
{\bf 2076299753393405312} &	5024526\,$^{X}$	      & 8.0142	&  18.520	& MS & -- & -- & -- & -- & -- & -- \\ 
{\bf 2076299688981553280} &	5024365\,$^{1S, 1L}$  & 7.4463	&  14.700	& MSTO & -- & -- & -- & -- & -- & -- \\ 
2076299001786815104 &	5024641\,$^{14L}$     & 5.6497	&  14.742	& MSTO	 & -- & -- & -- & -- & -- & -- \\ 
{\bf 2076298963119414912} &	5024756\,$^{X}$	      & 21.7471	&  17.979	& MS & -- & -- & -- & -- & -- & -- \\ 
{\bf 2076298417671212800} &	5024491\,$^{X}$	  & 4.0566	&  16.486	& MS & -- & -- & -- & -- & -- & -- \\ 
{\bf 2076298310287835776} &	4936866\,$^{X}$	  & 20.9493	    & 17.739	 &		--    &  --       &     --         &       --      & --\\
{\bf 2076582465319137792} &	5200072\,$^{X}$	  & 10.0885	&  18.497	& MS & -- & -- & -- & -- & -- & -- \\     
{\bf 2076581851136420608} & -- & 0.2074 & 20.413 & MS$^+$ & -- & -- & -- & -- & -- & --\\
2076396171125625984 &	5200422\,$^{14L}$ & 5.1932 &  14.950	& MS & -- & -- & -- & -- & -- & -- \\ 
{\bf 2076581675044794112} &	5112589\,$^{X}$	  & 7.3381	    &  17.476	& MS & -- & -- & -- & -- & -- & -- \\ 
{\bf 2076394075181506688} &	5112866\,$^{X}$	  & 0.4869	&  17.434	& MS & -- & -- & -- & -- & -- & -- \\ 
{\bf 2076394006454008320} &	5112675\,$^{X}$	  & 4.3123	&  16.720	    & MS & -- & -- & -- & -- & -- & -- \\ 
{\bf 2076394006461998336} &	5112698\,$^{X}$	  & 3.6742	&  15.321	& MS & -- & -- & -- & -- & -- & -- \\ 
{\bf 2076393903382776576} &	5112699\,$^{X}$	  & 2.3101	&  15.925	& MS & -- & -- & -- & -- & -- & -- \\ 
{\bf 2076393869023084288} &	5112979\,$^{11L}$ & 23.2479	&  15.034	& MS & -- & -- & -- & -- & -- & -- \\ 
{\bf 2076394002154821632} &	5112736\,$^{X}$	  & 21.9286	&  18.039	& MS & -- & -- & -- & -- & -- & -- \\ 
{\bf 2076393933435330688} &	5112777\,$^{X}$	  & 5.9733	&  17.445	& MS & -- & -- & -- & -- & -- & -- \\ 
{\bf 2076393899075931904} &	5112682\,$^{X}$	  & 8.7219	&  19.198	    & MS$^+$ & -- & -- & -- & -- & -- & -- \\ 
{\bf 2076393800303612032} &	5113045\,$^{X}$	  & 22.2631	&  15.054	& MS & -- & -- & -- & -- & -- & -- \\ 
{\bf 2076393903382776704} &	5112710\,$^{X}$	  & 19.6045	&  15.451	& MS & -- & -- & -- & -- & -- & -- \\ 
{\bf 2076393697224343296} &	5112781\,$^{X}$	  & 2.6031	&  18.105	& MS & -- & -- & -- & -- & -- & -- \\ 
{\bf 2076393869023085440} &	5113010\,$^{X}$	  & 1.4268	    &  15.357	& MS & -- & -- & -- & -- & -- & -- \\ 
{\bf 2076394178252904320} &	5113076\,$^{X}$	  & 13.3217	&  18.439	& MS & -- & -- & -- & -- & -- & -- \\ 
{\bf 2076393319267340288} &	5113478\,$^{X}$	  & 0.8781	&  16.560	& MS & -- & -- & -- & -- & -- & -- \\ 
2076393662865120256 &	5112908\,$^{X}$	  & 4.7402	&  16.980	& MS & MS & 5\,500(30) & 4.47(7) & 18(2) & -0.43(9) & 1\,$^*$ \\ 
{\bf 2076393731584598528} &	5112852\,$^{X}$	  & 4.5927	&  16.185	    & MS & -- & -- & -- & -- & -- & -- \\ 
{\bf 2076393628504856448} &	5024697\,$^{X}$	  & 3.3347	&  16.109	& MS & -- & -- & -- & -- & -- & -- \\ 
2076393692914036608 &	5112741\,$^{14L}$ & 17.6960	&  12.408	& RC & RC & 4\,814(107) &	2.64(17) & -38(5) & -0.08(10) & 3 \\ 
{\bf 2076393044389353984} &	5113099\,$^{X}$	  & 23.3089	&  17.201	& MS & -- & -- & -- & -- & -- & -- \\ 
2076393765944335104 &	5113001\,$^{14L}$ & 6.5676	&  14.687	& MSTO & -- & -- & -- & -- & -- & -- \\ 
2076393662865120256 &	5112908\,$^{X}$	  & 2.9060	&  16.980	& MS & -- & -- & -- & -- & -- & -- \\ 
{\bf 2076393284907576960} &	5113426\,$^{X}$	  & 7.3628	    &  18.081	    & MS & -- & -- & -- & -- & -- & -- \\ 
{\bf 2076393147460788864} &	5113295\,$^{11L}$ & 28.6162	&  16.112	& MS & RGB & 5\,420(400) & 3.5(5) & -34(6)	& 1.5(5) & 1\,$^*$ \\ 
{\bf 2076392838230907392} &	5024988\,$^{X}$	  & 5.4480	&  15.502	& MS & -- & -- & -- & -- & -- & -- \\ 
{\bf 2076392868287903872} &	5024774\,$^{X}$	  & 2.8742	&  17.551	& MS & -- & -- & -- & -- & -- & -- \\ 
2076392872590614784 &	5024751\,$^{X}$	  & 3.4953	&  17.904	& MS & -- & -- & -- & -- & -- & -- \\ 
{\bf 2076390295610350464} &	5113519\,$^{X}$	  & 10.9769	&  17.991	    & MS & -- & -- & -- & -- & -- & -- \\ 
{\bf 2076392730841568640} &	5025077\,$^{X}$	  & 21.7111	&  17.929	& MS & -- & -- & -- & -- & -- & -- \\ 
{\bf 2076392975669885568} &	5113165\,$^{X}$	  & 6.4769	&  16.297	& MS & -- & -- & -- & -- & -- & -- \\ 
{\bf 2076392799560963584} &	--	    	      & 20.2391	&  19.278	& MS$^+$ & -- & -- & -- & -- & -- & -- \\ 
{\bf 2076389711494716800} &	5025150\,$^{X}$	  & 6.3577	&  15.074	& MS & -- & -- & -- & -- & -- & -- \\ 
{\bf 2076389878983223808} &	5025470\,$^{X}$	  & 22.1879	&  18.216	& MS & -- & -- & -- & -- & -- & -- \\ 
{\bf 2076300066938729856} & 5112574\,$^{X}$	     & 1.6160  & 16.045	& MS & -- &	--	& --  & --  & --  & --  \\
{\bf 2076393800295734144} & 5113026\,$^{X}$      & 20.8330   & 18.069	& MS & -- &	--	& --  & --  & --  & --  \\
{\bf 2076298967427112448} & 5024801\,$^{X}$      & 1.8472   & 15.123	& MS & -- &	--	& --  & --  & --  & --  \\
{\bf 2076488049055863936} & -- & 6.8456& 19.194 & MS & -- &	--	& --  & --  & --  & --  \\
{\bf 2076299856472624128} & -- & 17.8793 & 18.135 & MS & -- &	--	& --  & --  & --  & --  \\
\hline\hline
\end{tabular}
}
\end{table*}
%%%%%%%%%%%%%%%%%%%%%%%%%%%%%%%%%%%%%%%%%%%%%%%%%%%%%%%%%%%%%%%%%%%%%

\subsection{Unclassified and unidentified variables}
In the case of 24 stars, we were unable to classify their variability types. The data are either not precise or it is not too clear if the shape of a light curve remains stable over time. The latter argument is essential to distinguish between binaries and rotational variables. An example of a light curve in an unclassified group is presented in the bottom panel of Fig.\,3. We identified five variables to be cluster members, 18 non-members and one with membership undetermined, and they are listed in Table\,4,\,9 and 10, respectively. All field stars, four members and the one with no membership, are new variables detected in this work. The newly discovered variable KIC\,5023913 exhibits low frequency pulsations, similar to solar-like ones, and is positioned near the MS on the CMD. However, it cannot be classified as a solar-like pulsator due to the limited sampling of the \kep\ long cadence data, which is insufficient to detect MS pulsation occurring beyond the Nyquist limit. KIC\,5112843 shows two close frequencies with high amplitudes, which translates into beating in a light curve. The star may be a binary system, as discussed by Guzik \etl(2023), however it does not fit into any of our four binary classes. We leave such cases for future investigation. The location of the cluster members in the CMD is shown in the bottom panel of Fig.\,5.

We found signals in amplitude spectra, associated with optical counterparts in the Pan-STARRS (Chambers \etl 2016, Flewelling \etl 2020) survey that do not have \gaia\ designations. Only one star has a custom designation and was reported in the past. The remainder of the stars are given by coordinates. We classified the signal to a proper variability type and estimated its period. These stars were included in the counts of the variability types discussed in the previous subsections. We show the list of these unidentified objects in Table\,10. Since the stars are not listed in the \gaia\ catalog, we are unable to estimate their membership.

%\begin{center}
%\input{table4.tex}
%\end{center}
%%%%%%%%%%%%%%%%%%%%%%%%%%%%%%%%%%%%%%%
%Table of unclassified variables IN the CLUSTER
\begin{table*}
\caption{List of cluster members with unclassified variability. See caption of Table\,1 for explanation.}
\label{tab:cluster_unclassified}
\centering
\resizebox{\columnwidth}{!}{%
\begin{tabular}{clrccclllll}
\hline\hline
\multicolumn{1}{c}{\multirow{2}{*}{\gaia\ DR3}} & \multicolumn{1}{c}{\multirow{2}{*}{KIC}} & \multicolumn{1}{c}{Period} & \multicolumn{1}{c}{G} & \multirow{2}{*}{CMD} & \multirow{2}{*}{HRD} & \multicolumn{1}{c}{T$_{\rm eff}$} & \multicolumn{1}{c}{\multirow{2}{*}{\logg}} & \multicolumn{1}{c}{RV} & \multicolumn{1}{c}{\multirow{2}{*}{[Fe/H]}} & \multicolumn{1}{c}{\multirow{2}{*}{Ref}}\\
&& \multicolumn{1}{c}{[days]} & \multicolumn{1}{c}{[mag]} &&& \multicolumn{1}{c}{[K]} && \multicolumn{1}{c}{[km/s]} &&\\
\hline\hline
{\bf 2076488353993641344} & 5111986\,$^{X}$ & 0.6185    & 18.188 & MS & -- & -- & -- & -- & -- & -- \\ 
{\bf 2076299306716922752} & 5024211\,$^{X}$      & 1.1752     & 18.711	& MS & -- & -- & -- & -- & -- & -- \\ 
%{\bf 353-2076393834655409536} & 5112918\,$^{X}$      & 1.0003     & 18.622	& MS$^+$& -- \\
{\bf 2076393765943855744} & 5112994\,$^{X}$      & 1.0031     & 16.729	& MS & -- & -- & -- & -- & -- & -- \\ 
%391-2076393731584598400 & 5112843\,$^{X}$      & 0.1743           & 15.422	& MS & RGB\\
{\bf 2076299414103588224} &	5023913\,$^{X}$       & 18.3197 & 15.274	& MS & --	& 	--		&    --       &        --   &    --       & --   \\
2076393731584598400 & 5112843\,$^{X}$      & 0.1743           & 15.422	& MS & RGB & 5\,600(50)	& 4.37(5) &  42(1) & -1.4(1) & 2*\\
\hline\hline
\end{tabular}
}
\flushleft
\end{table*}
%%%%%%%%%%%%%%%%%%%%%%%%%%%%%%%%%%%%%%%%%%%%%%%%%

\section{The distance and age estimation}
We downloaded a grid of isochrones given in the \gaia\ photometric system from the MIST project (Choi \etl 2016, Dotter\,2016). The current version of MIST is 1.2. The MESA version 7503 was employed to calculate isochrones. We selected V/V$_{\rm crit}$\,=\,0. The grid covers age in a logarithmic scale between 9.1 and 9.6 with a step of 0.01 and the iron content [Fe/H] from -0.1 to +0.1 with a step of 0.01.

For the fit, we only kept the MS, RGB and RC stars. However, in case of binary systems, the observed magnitude may include the flux contribution from all companions and not a single star, which shifts the position of a system in the CMD. To avoid biased magnitudes, we excluded outlying stars by considering their positions in the CMD as uncertain. We included magnitude uncertainties as weights in the fit, which prevented the MS targets from over-fitting. The RC and RGB targets, even though less numerous, are brighter, and hence remain significant in the fit.

The MIST synthetic isochrones are given in absolute magnitudes, and we selected no extinction. We applied a shift (m-M) in the \gaia\ G magnitude and B$_{\rm p}$-R$_{\rm p}$ color to account for the extinction. The isochrones that fit the best are defined by two consecutive values of the age and, for each age, the same three [Fe/H] values. This solution is quite common in astrophysical modelling. The fitting shows a degeneration of solutions in which a pair of different age and [Fe/H] values compensate by providing similar quality fits. We have determined the age to be 2.54(3)\,Gyr and [Fe/H] to be -0.01(2). The apparent distance modulus\,(m-M) equals 12.20 and E(B$_{\rm p}$-R$_{\rm p}$) to be 0.195(10). We show the best isochrone fits in Fig.\,7. 

For the \gaia\ DR3 cluster members E(B$_{\rm p}$-R$_{\rm p}$) and interstellar extinction in the G band, the A$_{\rm G}$ form a linear relation A$_{\rm G}$\,=\,2$\cdot$E(B$_{\rm p}$-R$_{\rm p}$). Averaging the parameters from the best models we obtain E(B$_{\rm p}$-R$_{\rm p}$)\,=\,0.194\,mag, which corresponds to A$_{\rm G}$\,=\,0.388\,mag. Subtracting A$_{\rm G}$ from (m-M) we find the true distance modulus of 11.812, which gives the distance to the cluster of 2.3\,kpc.

%\begin{center}
%\input{table5.tex}
%\end{center}
%FIELD BINARIES
\begin{table*}
\caption{The list of binary stars in the field that are not cluster members or no memberships derived. See caption of Table\,1 for explanation. Targets with no astrometry, hence no membership established, are denoted in {\it italic}.}
\label{tab:field_binaries}
\centering
\resizebox{\columnwidth}{!}{
\begin{tabular}{clllclllll}
\hline\hline
\multicolumn{1}{c}{\multirow{2}{*}{\gaia\ DR3}} & \multicolumn{1}{c}{\multirow{2}{*}{KIC}} & \multicolumn{1}{c}{P$_{\rm orb}$} & \multicolumn{1}{c}{T$_0$} & \multicolumn{1}{c}{G} & \multicolumn{1}{c}{T$_{\rm eff}$} & \multicolumn{1}{c}{\multirow{2}{*}{\logg}} & \multicolumn{1}{c}{RV} & \multicolumn{1}{c}{\multirow{2}{*}{[Fe/H]}} & \multicolumn{1}{c}{\multirow{2}{*}{Ref}}\\
&& \multicolumn{1}{c}{[days]} & \multicolumn{1}{c}{[BJD]} & \multicolumn{1}{c}{[mag]} & \multicolumn{1}{c}{[K]} && \multicolumn{1}{c}{[km/s]} &&\\
\hline\hline
\multicolumn{10}{c}{eclipsing}\\
\textbf{2076487671098665472} &		 --    		& 225.27602919(47)$^e$& 2\,455\,003.738545(5)    & 19.596	&  -- & -- & -- & -- & -- \\
\textbf{2076299757701057920} &		--     		& 38.857349(27)$^e$	  & 2\,454\,997.2392(6)      & 19.489	& -- &  -- & -- & -- & -- \\
\textbf{2076298898703420416} &		 --    		& 0.42517143(28)  & 2\,454\,971.20746(53)    & 20.682	& -- &  -- & -- & -- & -- \\
\textbf{2076297657449418880} &	5023984\,$^{X}$	& single-eclipse  & 2\,455\,008.4490(24)     & 18.400	& -- &  -- & -- & -- & -- \\
\textbf{2076298417671214464} &	5024527\,$^{X}$	&30.564505(25)$^e$	  & 2\,454\,986.35489(57)    & 15.876	& 5\,910(200) & 4.736(22) & -75(14) & -0.68(13) &  4\,$^*$     \\
\textbf{2076583294245566720} &		 --    		& 2.8083498(28)	  & 2\,454\,967.2876(7)      & 20.461	& -- &  -- & -- & -- & -- \\
2076583839713435776 &	5200364\,$^{X}$	& 0.206879465(16) & 2\,454\,964.58302(7)     & 19.420	& -- &  -- & -- & -- & -- \\
2076394590569930240 &		--     		& 0.8907258(10)	  & 2\,454\,965.42331(9)     & 20.064	& -- &  -- & -- & -- & -- \\
2076394178252936704 &	5113146\,$^{1S/3L}$	& 18.789711(29)	  & 2\,456\,216.94558(19)    & 19.193	& -- &  -- & -- & -- & -- \\
2076390123803775104 &	5025294\,$^{14L}$	& 5.462678(18)$^e$	  & 2\,455\,021.8935(23)     & 13.246	& -- &  -- & -- & -- & -- \\
\textbf{2076390119501429120} &	5025349\,$^{X}$	& 50.71339(6)$^e$	  & 2\,454\,967.2679(10)     & 18.897	& -- &  -- & -- & -- & -- \\
\textbf{2076298692545230848} &		--     		& single-eclipse  & 2\,455\,156.9573(25)     & 19.816	& -- &  -- & -- & -- & -- \\

\hline
\multicolumn{10}{c}{active eclipsing}\\
\textbf{2076487495009192064} &	5023833\,$^{X}$	& 1.5995032(46)	  & 2\,454\,965.0930(26)     & 19.173	&  -- & -- & -- & -- & -- \\
2076487495000121216 &	5023832\,$^{X}$	& 0.357122835(39) & 2\,454\,964.69840(9)     & 19.523	& -- &  -- & -- & -- & -- \\
\textbf{2076487293141296640} &	5023901\,$^{X}$	& 318.7285(20)$^e$	  & 2\,455\,009.7943(50)     & 19.557	& -- &  -- & -- & -- & -- \\
\textbf{2076487636738867328} &	--	     		& 0.4169454(7)	  & 2\,454\,965.2280(24)     & 19.745	& -- &  -- & -- & -- & -- \\
\textbf{2076394006453983488} &	--	     		& 3.0224820(6)	  & 2\,454\,964.98930(15)    & 20.519	& -- &  -- & -- & -- & --\\
2076487254481737728 &	5024077\,$^{X}$	& 0.932208699(49) & 2\,454\,965.28509(44)    & 13.984   & 16\,860(230)&	5.79(16)&	6(5)&	-4(2) & 4\,$^*$  \\
2076299276660357760 &	5024146\,$^{X}$	& 0.787121288(48) & 2\,454\,964.73835(41)    & 15.588	& 5\,810(200)	& 4.03(1)	&	  -13(9)     & -0.45(27) &  4\,$^*$ \\
2076393284907582464 &	5113407\,$^{X}$	& 13.826433(5)$^e$	  & 2\,454\,975.78903(29)    & 18.049	& -- &  -- & -- & -- & -- \\
2076393044381456896 &	5113053\,$^{1S/4L}$	& 3.1850900(12)	  & 2\,455\,466.35322(11)    & 20.230	& -- &  -- & -- & -- & -- \\
\textit{\textbf{2076394659289386880}} &	5200325\,$^{X}$	& 5.3159294(28)$^e$	  & 2\,454\,969.74506(27)    & 17.393	& 5\,380(200) & 5.04(8) &	-40(13)	& -0.81(15) & 4\,$^*$ \\

\hline
\multicolumn{10}{c}{contact}\\
2076489182934321920 &	5199669\,$^{X}$	& 0.758956896(42) & 2\,454\,964.903723(46)   & 15.226	&  -- & -- & -- & -- & -- \\
2076488285274150784 &	5111817\,$^{X}$	& 0.28994151(7)	  & 2\,454\,964.73887(21)    & 19.638	&  -- & -- & -- & -- & -- \\
2076487739818143360 &	--	     	    & 0.28136943(9)	  & 2\,454\,964.71482(27)	 & 20.462	&  -- & -- & -- & -- & -- \\
2076298452030939392 &	5024283\,$^{X}$	& 0.33846178(5)	  & 2\,454\,964.58443(13)    & 18.044	& -- & -- & -- & -- & --\\
\textbf{2076298314587607040} &	--	     	    & 0.223753315(41) & 2\,454\,964.56924(15)	 & 20.089		&  -- & -- & -- & -- & -- \\
2076394521858125696 &	5112917\,$^{X}$	& 0.29314610(6)	  & 2\,454\,964.66198(17)    & 16.878   & -- & -- & -- & -- &  -- \\
\textbf{2076393697224860544} &	5112791\,$^{X}$	& 0.9816957(5)	  & 2\,454\,964.78683(42)	 & 14.669		&  -- & -- & -- & -- & -- \\
2076393216188094336 &	5113461\,$^{X}$	& 0.275128153(42) & 2\,454\,964.73810(13)    & 19.632		&  -- & -- & -- & -- & -- \\
2076389745856651136 &	5025261\,$^{15L}$	& 2.171411(7)	  & 2\,454\,966.95452(29)	 & 12.296		&  -- & -- & -- & -- & -- \\
\textbf{2076487464939938816} &	5023779\,$^{X}$	& 0.1963          &	    --    	         & 17.826	& 	--		&     --      &    --      &  --       & --  \\
%4	& 	                  &                 & 0.27341359(26)  & 2454964.8998(7)      &              &           &           &           &          &          \\
\hline
\multicolumn{10}{c}{outbursting}\\
2076392906942365568 &       --       & -- & -- & 20.427 	&  -- & -- & -- & -- & -- \\
\textbf{2076392902640277632} & 5024812\,$^{X}$& -- & -- & 16.431	&  -- & -- & -- & -- & -- \\

%2	& 	                  &                 & 0.2342          &                      &              &           &           &           &          &          \\
%76	& 	                  &                 & 0.4093          &                      &              &           &           &           &          &          \\
%88	& 	                  &                 & 0.3660          &                      &              &           &           &           &          &          \\
%165	& 	                  &                 & 0.2080          &                      &              &           &           &           &          &          \\
%296	& 	                  & 				& 0.8824          &                      &              &           &           &           &          &          \\

\hline\hline
\end{tabular}
}
\end{table*}

%FIELD PULSATORS
\begin{table*}
\caption{The list of pulsators in the field that are not cluster members or no memberships derived. See captions of Table\,1 and 5 for explanation.}
\label{tab:Pulsators_field}
\centering
\resizebox{\columnwidth}{!}{%
\begin{tabular}{clclllll}
\hline\hline
\multicolumn{1}{c}{\multirow{2}{*}{\gaia\ DR3}} & \multicolumn{1}{c}{\multirow{2}{*}{KIC}} & \multicolumn{1}{c}{G} & \multicolumn{1}{c}{T$_{\rm eff}$} & \multicolumn{1}{c}{\multirow{2}{*}{\logg}} & \multicolumn{1}{c}{RV} & \multicolumn{1}{c}{\multirow{2}{*}{[Fe/H]}} & \multicolumn{1}{c}{\multirow{2}{*}{Ref}}\\
&& \multicolumn{1}{c}{[mag]} & \multicolumn{1}{c}{[K]} && \multicolumn{1}{c}{[km/s]} &&\\
\hline\hline
\multicolumn{8}{c}{solar-like}\\
2076489251653787904 &	5199605\,$^{2L}$       & 11.496	 & 4\,250(50)   &	3.00(5) &	0.2(1)	 & -0.19(3)  & 5\,$^*$ \\
2076488495739440128 &	5111767\,$^{6L}$       & 13.641	 &	4\,524(130) &	2.66(20) & -49(3)	 & 0.20(12)   & 3 \\
2076488804977149568 &	5111987\,$^{14L}$       & 14.949 &	--	      &	   --      &     --      &       --         & --\\
\textbf{2076488117782408192} &	5112211\,$^{X}$	       & 14.424	 &	--	      &  --        &     --      &        --        & --\\
\textbf{2076488014703171328} &	5112103\,$^{X}$	       & 14.804	 &	--	      &	   --      &     --      &        --        & -- \\
\textbf{2076488598818838400} &	5111863\,$^{X}$	       & 14.886	 &	--	      &   --       &     --      &        --        &  --\\
\textbf{2076488873696651264} &	5112156\,$^{X}$	       & 15.403	 &	--	      &	 --        &     --      &        --        & --\\
\textbf{2076487945983698432} &	5112169\,$^{1S, 15L}$  & 10.864	 &	6\,283(11)  & 3.927(18)& 39(4)	 & 0.155(9)   & 3 \\
2076300135658208384 &	5112558\,$^{10L}$      & 14.112	 &	--	      &     --     &     --      &       --         & --\\
\textbf{2076299963859543680} &	5024196\,$^{X}$	       & 13.291	 &	--	      &     --     &     --      &       --         & --\\
\textbf{2076299895139970816} &	5024238\,$^{X}$	       & 13.612	 & -- & -- & --	 & -- & -- \\
\textbf{2076299585902315264} &	5024100\,$^{X}$	       & 13.609	 &	--	      &    --      &    --       &       --         & --\\ 
\textbf{2076299688981538944} &	5024290\,$^{X}$	       & 15.275	 &	--	      &	   --         &   --        &    --            & --\\
\textbf{2076298967427099136} &	5024773\,$^{3L}$       & 14.325	 &	--	      &	    --     &    --       &       --         & --\\ 
\textbf{2076298761268662016} &	5024804\,$^{X}$	       & 15.290	 &	--	      &	    --     &     --      &       --         & --\\ 
2076299139225684096 &	5024043\,$^{15L}$      & 12.779	 & 4\,762(92)   &	2.70(11) &	2.82(14) &	-0.10(3)      & 5   \\
\textbf{2076298280232974976} &	4936972\,$^{X}$	       & 13.597	 &	--	      &   --       &     --      &       --         & --\\
2076296837123206912 &	4936825\,$^{4L}$       & 13.144	 &	--	      &  --         &    --       &       --         & --\\
2076582916298933888 &	5199930\,$^{5L}$       & 11.893	 & 4\,260(50)   & 1.94(5)	 & 0.2(3) & -0.25(9)  & 5\,$^*$\\
\textbf{2076582503982578944} &	5200146\,$^{1L}$       & 11.611	 & 4\,069(144)  & 2.15(23) & -20(3)    & -0.22(14)  & 3 \\
\textbf{2076582297823683712} &	5200223\,$^{X}$	       & 14.140	 & 	--	      &    --      &   --        &         --       & --\\	
\textbf{2076582366543185664} &	5200359\,$^{X}$	       & 13.118	 & 	--	      &    --      &    --       &     --           & --\\
\textbf{2076394693656870400} &	5200392\,$^{X}$	       & 14.772	 & 	--	      &    --      &    --       &      --          & --\\	
\textbf{2076582362237411712} &	5200367\,$^{X}$	       & 13.063	 & 	--	      &    --      &    --       &       --         & --\\
\textbf{2076581919866487808} &	5112421\,$^{X}$	       & 15.505	 &    --        &   --        &   --       &      --          & --\\	
\textbf{2076582263463928064} &	5200143\,$^{X}$	       & 15.564	 & 	--	      &   --        &  --         &        --        & --\\ 	
\textbf{2076394418778961024} &	5113246\,$^{X}$	       & 14.720	 & 	--	      &  --          &    --       &        --        & --\\	
\textbf{2076393899075590528} &	5112672\,$^{X}$	       & 16.690	 & 	--	      &  --        &   --        &           --     & --\\	
\textbf{2076393594145243136} &	5113371\,$^{X}$	       & 14.452	 & 	--	      &  --          &    --       &       --         & --\\	
\textbf{2076392700791972992} &	5025162\,$^{15L}$      & 12.623	 & 	 --	      &  --        &  --         &           --     & --\\	
2076392769511450368 &	5025101\,$^{2L}$       & 13.003	 & 4\,404(74)   & 2.73(12) &	-60(4)	 & 0.10(7)   & 3 \\	
\textbf{2076389711494720640} &	5025168\,$^{X}$	       & 14.392	 & --		      &     --     &  --         &         --       & --\\	
\textbf{2076392563352990208} &	5025060\,$^{X}$	       & 14.124	 & --		      &    --      &   --        &          --      & --\\	
2076392528993243008 &	5025021\,$^{6L}$       & 12.842	 & 4\,675(92)   & 2.65(11) &	38.86(46)&	0.100(28)     & 5 \\

\hline
\multicolumn{8}{c}{$\delta$-scuti}\\
\textbf{2076298211512787328} &	5024570\,$^{X}$	       & 16.935	 & 7\,760(60)&	3.926(24) & 27.1(8) & -0.57(7)  & 2\,$^*$   \\
2076395896247722496 &	5200521\,$^{1S,15L}$   & 11.182	 & 5\,810(32) & 3.535(53) & -10(6) & 0.276(30) & 3 \\

\hline
\multicolumn{8}{c}{$\gamma$\,Doradus}\\
\textbf{2076487632439073408} &	5024090\,$^{X}$	       & 19.103	 &	--	   &	 --        &   --        &     --   &  --     \\
\textbf{2076582602761996672} &	5200181\,$^{X}$	       & 17.216	 & 7\,150(50)&	4.80(5) & -5.6(7) & -0.46(9)  & 2\,$^*$     \\
\textbf{2076394315699722368} &	5113250\,$^{14L}$      & 14.439	 & 5\,430(50)&	3.0(1)& -51(2)  & -0.65(7) & 2\,$^*$      \\
2076390192533257216 &	5025464\,$^{15L}$      & 13.122	 & 7\,182(30)&	4.078(50)& -7(9)   & -0.243(28) & 3 \\
\textbf{2076392627762286336} &	5025047\,$^{X}$	       & 18.250	 & 6\,930(40)&	3.35(7)& 45(5)   & -0.40(7) & 2\,$^*$       \\
2076389642775246080 &	5025234\,$^{15L}$      & 13.285	 & 		--   &       --      &   --        &     --     &  -- \\

\hline
\multicolumn{8}{c}{semi-regular}\\
2076487808546982784 & 5112438\,$^{15L}$ & 10.602 & 3\,715(69)	& 0.94(8)	& -38.40(34) &	-0.400(29) & 5\\
2076298829988110976 & 5024699\,$^{15L}$ & 11.600 & 3\,500	& 1	& -45.9053(69)	& -0.5 & 5\\
2076392632072463232 & 5025003\,$^{15L}$ & 8.011  & 		--   &       --      &   --        &     --     &  -- \\

\hline\hline
\end{tabular}
}
\end{table*}

%Field Rotational part 1
\begin{table*}
\caption{The list of rotational variables that are not cluster members or no memberships derived. See captions of Table\,1 and 5 for explanation.}
\label{tab:Rotational_field1}
\centering
\resizebox{\columnwidth}{320pt}{%
\begin{tabular}{clrclllll}
\hline\hline
\multicolumn{1}{c}{\multirow{2}{*}{\gaia\ DR3}} & \multicolumn{1}{c}{\multirow{2}{*}{KIC}} & \multicolumn{1}{c}{Period} & \multicolumn{1}{c}{G} & \multicolumn{1}{c}{T$_{\rm eff}$} & \multicolumn{1}{c}{\multirow{2}{*}{\logg}} & \multicolumn{1}{c}{RV} & \multicolumn{1}{c}{\multirow{2}{*}{[Fe/H]}} & \multicolumn{1}{c}{\multirow{2}{*}{Ref}}\\
&& \multicolumn{1}{c}{[days]} & \multicolumn{1}{c}{[mag]} & \multicolumn{1}{c}{[K]} && \multicolumn{1}{c}{[km/s]} &&\\
\hline\hline
2076489006828963200 &	     --	          & 4.2194	    & 20.192	     &	--	    &    --    &	--	         &     --      & --\\
\textbf{2076491656835399552} &	5111748\,$^{X}$	  & 2.2534	    & 14.809	     &	--	    &    --    &    --          &      --      & --\\
\textbf{2076488972469006720} &	   --  	          & 4.3523	    & 19.439	 &	--	    &    --    &    --          &      --      & --\\	       
\textbf{2076491588108521600} &	5111658\,$^{X}$	  & 11.2363	    & 18.431	 &	--	    &    --    &    --          &      --      & --\\
2076489212987407872 &	5199564\,$^{X}$	  & 3.4612	    & 18.809	 &	--	    &    --    &    --          &      --      & --\\		
\textbf{2076488976768870656} &	     --	          & 4.0641	    & 19.688	 &	--	    &    --    &    --          &      --      & --\\	
\textbf{2076491656828123904} &	     --	          & 11.6758	    & 19.207	 &	--	    &    --    &    --          &      --      & --\\
\textbf{2076489109908182400} &	     --	          & 1.8813	    & 20.143	 &	--	    &    --    &    --          &      --      & --\\
\textbf{2076489079855048320} &	5111878\,$^{X}$	  & 5.2625	    & 16.775	 &	--	    &    --    &    --          &      --      & --\\
2076492030485722880 &	5111830\,$^{X}$	  & 3.1613	    & 18.914	 &	--	    &    --    &    --          &      --      & --\\
\textbf{2076488835030026752} &	5112025\,$^{X}$	  & 6.1208	    & 18.934	 &	--	    &    --    &    --          &      --      & --\\
\textbf{2076491656835403776} &	5111760\,$^{X}$	  & 6.7188	    & 15.758     &	--	    &    --    &    --          &      --      & --\\
\textbf{2076491519396420864} &	5111722\,$^{X}$	  & 2.3987	    & 19.046	 &	--	    &    --    &    --          &      --      & --\\		
\textbf{2076491618168829184} &	5111737\,$^{X}$	  & 2.1654	    & 16.894	 &	--	    &    --    &    --          &      --      & --\\
2076488873696743936 &	5112157\,$^{14L}$ & 27.9330	    & 15.165	 &	--	    &    --    &    --          &      --      & --\\		
2076488873696744064 &	5112117\,$^{14L}$ & 13.6986	    & 14.966	 &	--	    &    --    &    --          &      --      & --\\		
\textbf{2076488835038990976} &	5112066\,$^{X}$	  & 11.1315	    & 18.715	 &	--	    &    --    &    --          &      --      & --\\
2076488633178431232 &	5111890\,$^{14L}$ & 8.2095	    & 13.851	 &	--	    &    --    &    --          &      --      & --\\		
2076488427020146176 &	5111932\,$^{14L}$ & 1.4345  	& 12.215	 & 6\,556(16)	& 4.216(27)&	-19(5)	 & -0.133(15) & 3 \\
\textbf{2076488628871828992} &	5111870\,$^{X}$	  & 5.1734	    & 18.803     &	--	    &    --     &      --        &      --       & --\\
\textbf{2076488152135136256} &	     --	          & 0.9692	    & 20.024	 &	--	    &    --     &      --        &      --       & --\\
\textbf{2076488113482225920} &	5112203\,$^{X}$	  & 24.0132	    & 17.997	 &	--	    &    --     &      --        &      --       & --\\
\textbf{2076488731950773760} &	5112036\,$^{X}$	  & 4.4587	    & 17.976	 &	--	    &    --     &      --        &      --       & --\\
2076488495732070016 &	5111802\,$^{X}$	  & 14.2002	    & 19.035	 &	--	    &    --     &      --        &      --       & --\\
\textbf{2076488461379688448} &	5111755\,$^{X}$	  & 26.8169	    & 12.863	 &	--	    &    --     &      --        &      --       & --\\		
\textbf{2076488014703180416} &	5112144\,$^{X}$	  & 26.8614	    & 17.006	 &	--	    &    --     &      --        &      --       & --\\
\textbf{2076488491432593024} &	5111774\,$^{X}$	  & 20.2080     & 17.369	 &	--	    &    --     &      --        &      --       & --\\
\textbf{2076488427019997568} &	5111912\,$^{X}$	  & 19.7522	    & 16.803	 &	--	    &    --     &      --        &      --       & --\\
\textbf{2076488044756006528} &	5112226\,$^{X}$	  & 3.1371	    & 19.372	 &	--	    &    --     &      --        &      --       & --\\
\textbf{2076487945983706880} &	5112224\,$^{X}$	  & 7.7540	    & 16.816	 &	--	    &    --     &      --        &      --       & --\\
\textbf{2076487602386283520} &	5112061\,$^{X}$	& 0.3031  & 16.601	& -- &  -- & -- & -- & -- \\
\textbf{2076487872957580544} &	5112420\,$^{X}$	  & 10.1352	    & 18.165	 &	--	    &    --     &      --        &      --       & --\\
2076581644988572160 &	5112508\,$^{3L}$  & 0.7502	    & 15.189	 &	--	    &    --     &      --        &      --       & --\\		
\textbf{2076487907323097600} &	5112158\,$^{X}$	  & 1.2659	    & 18.122	 &	--	    &    --     &      --        &      --       & --\\
\textbf{2076487705465511936} &	5024046\,$^{X}$	  & 7.2570	    & 15.464	 &	--	    &    --     &      --        &      --       & --\\
\textbf{2076487357561147264} &	5023750\,$^{X}$	  & 7.2731	    & 18.786	 &	--		&    --     &      --        &      --       & --\\
2076300135658196864 &	5112483\,$^{14L}$ & 62.8479	    & 11.401	 &	--	    &    --     &      --        &      --       & --\\
2076487636746036736 &	5024079\,$^{14L}$ & 20.000      & 15.152	 &	--	    &    --     &      --        &      --       & --\\
\textbf{2076487739825262976} &	5112228\,$^{14L}$ & 21.0838	    & 14.138	 & 5\,653(133)&4.67(0.22)& -3(6)	     & -0.02(13) & 3\\
\textbf{2076581640685054208} &	5112520\,$^{X}$	  & 15.1677	    & 17.561	 &	--    &   --      &     --         &    -         & --\\
\textbf{2076300032574913920} &	     --	          & 0.5615	    & 20.904	 &	--    &   --      &     --         &    -         & --\\
\textbf{2076487258788897792} &	5024021\,$^{X}$	  & 1.0110	    & 19.549	 &	--    &   --      &     --         &    -         & --\\
\textbf{2076487293141276672} &	5023904\,$^{X}$	  & 32.5945	    & 20.454	 &	--    &   --      &     --         &    -         & --\\
\textbf{2076487190069384960} &	5023844\,$^{X}$	  & 5.4594	    & 19.614	 &	--    &   --      &     --         &    -         & --\\
\textbf{2076487190061988608} &	     --	          & 0.1887	    & 20.733	 &	--    &   --      &     --         &    -         & --\\
2076487151403028224 &	     --	          & 3.5949	    & 19.612	 &	--    &   --      &     --         &    -         & --\\
\textbf{2076299860780304512} &	5024541\,$^{X}$	  & 7.4085	    & 15.600	 &	--    &   --      &     --         &    -         & --\\
\textbf{2076299895135780224} &	5024204\,$^{X}$	  & 0.5714	    & 19.901	 &	--    &   --      &     --         &    -         & --\\
\textbf{2076299551542571264} &	5024070\,$^{X}$	  & 16.2968     & 13.604	 &	3\,883(92)& 0.73(11)&	-84.49(29)   & -0.500(31)  & 5 \\
2076299890832483584 &	5024233\,$^{X}$	  & 6.4642	    & 18.150	 &		--	&  --       &     --         &    --         & --\\	
2076299895139968000 &	5024215\,$^{14L}$ & 4.4271	    & 13.396	 & 6\,491(42)	& 4.12(7) & -27(7)       & 0.01(40) & 3 \\
\textbf{2076299860776267136} &	5024529\,$^{X}$	  & 12.5097	    & 17.884	 &		--    &  --       &     --         &       --      & --\\
2076299895135748864 &	5024202\,$^{11L}$ & 2.9086    	& 17.612	 &		--    &  --       &     --         &       --      & --\\
2076299757697024128 &	     	--          & 6.3272	    & 18.758	 &		--    &  --       &     --         &       --      & --\\
\textbf{2076299654617645056} &	     	--          & 1.2605	    & 19.506	 &		--    &  --       &     --         &       --      & --\\
\textbf{2076299684674005504} &	5024300\,$^{X}$	  & 6.6880	    & 17.767	 &		--    &  --       &     --         &       --      & --\\
2076299345384113536 &	5023985\,$^{15L}$ & 6.5381   	& 13.453	 &		--    &  --       &     --         &       --      & --\\
\textbf{2076299482818776704} &	     	--          & 10.6186	    & 19.999	 &		--    &  --       &     --         &       --      & --\\
\textbf{2076299753396856576} &	5024581\,$^{X}$	  & 3.9925	    & 18.570	 &		--    &  --       &     --         &       --      & --\\
2076299207945181824 &	5024254\,$^{14L}$ & 34.9997	        & 13.878	 &		--    &  --       &     --         &       --      & --\\
\textbf{2076299242304936192} &	5024335\,$^{X}$	  & 6.0531	    & 15.292	 &		--    &  --       &     --         &       --      & --\\
\textbf{2076299139225682688} &	5024055\,$^{X}$	  & 7.1974	    & 15.755	 &		--    &  --       &     --         &       --      & --\\
\textbf{2076298726908910336} &	5024794\,$^{X}$	  & 28.2478	    & 15.957     &		--    &  --       &     --         &       --      & --\\
\textbf{2076298722601233792} &	5024811\,$^{X}$	  & 2.2620	    & 15.734	 &		--    &  --       &     --         &       --      & --\\
\textbf{2076298589469925632} &	5024647\,$^{X}$	  & 8.3029	    & 15.259	 &		--    &  --       &     --         &       --      & --\\
\textbf{2076298795628342912} &	5024515\,$^{X}$	  & 4.7317	    & 16.948	 &		--    &  --       &     --         &       --      & --\\
2076298310284438656 &	4936870\,$^{X}$	  & 3.8084	    & 17.681     &		--    &  --       &     --         &       --      & --\\
\textbf{2076298314587586304} &	4936842\,$^{X}$	  & 0.7882	    & 16.477	 &		--    &  --       &     --         &       --      & --\\
\textbf{2076298486386442112} &	5024389\,$^{X}$	  & 2.3891	    & 20.433	 &		--    &  --       &     --         &       --      & --\\
2076582607061342336 &	5200184\,$^{13L}$ & 7.5913  	& 15.584	 &		--    &  --       &     --         &       --      & --\\
\textbf{2076582534038273920} &	5200034\,$^{X}$	  & 3.8284	    & 19.555	 &		--	&     --    &        --      &       --      & --\\
\hline\hline
\end{tabular}
}
\end{table*}

%Field Rotational
\begin{table*}
\caption{A continuation of the list of rotational variables that are not cluster members or no memberships derived. See captions of Table\,1 and 5 for explanation.}
\label{tab:rotational_field2}
\centering
\resizebox{\columnwidth}{320pt}{%
\begin{tabular}{clrclllll}
\hline\hline
\multicolumn{1}{c}{\multirow{2}{*}{\gaia\ DR3}} & \multicolumn{1}{c}{\multirow{2}{*}{KIC}} & \multicolumn{1}{c}{Period} & \multicolumn{1}{c}{G} & \multicolumn{1}{c}{T$_{\rm eff}$} & \multicolumn{1}{c}{\multirow{2}{*}{\logg}} & \multicolumn{1}{c}{RV} & \multicolumn{1}{c}{\multirow{2}{*}{[Fe/H]}} & \multicolumn{1}{c}{\multirow{2}{*}{Ref}}\\
&& \multicolumn{1}{c}{[days]} & \multicolumn{1}{c}{[mag]} & \multicolumn{1}{c}{[K]} && \multicolumn{1}{c}{[km/s]} &&\\
\hline\hline
\textbf{2076582946355475968} &	     	--          & 13.7111	    & 19.060	 &	  --	    &  --       &         --     &     --       &--\\
\textbf{2076582843276554240} &	     	--          & 7.8372	    & 20.120	 &	  --		&  --       &         --     &     --       &--\\
2076582607061335936 &	5200185\,$^{11L}$ & 3.7272	    & 15.565	 &	  --		&  --       &         --     &     --       &--\\
2076584118889853696 &	5200273\,$^{X}$	  & 6.5349	    & 16.913	 &	4\,390(60)	& 3.25(3)	& -46(1)	& -0.18(13) & 1\,$^*$ \\
\textbf{2076583908425657472} &	5200352\,$^{X}$	  & 6.6710	    & 18.897	 &	  --		&  --       &         --     &     --       &--\\
2076582087361668608 &	5199996\,$^{X}$	  & 6.4674	    & 19.317	 &	  --	    &  --       &         --     &     --       &--\\
\textbf{2076582538341833344} &	5200036\,$^{X}$	  & 21.9286	    & 16.748	 &	  --	    &  --       &         --     &     --       &--\\
2076582018642530432 &	5112402\,$^{10L}$ & 4.8263	    & 17.044	 &	  --	    &  --       &         --     &     --       &--\\
\textbf{2076582362237425920} &	5200342\,$^{X}$	  & 5.8353	    & 18.743	 &	  --		&  --       &         --     &     --       &--\\
\textbf{2076582465319431296} &	5200038\,$^{11L}$ & 2.6316	    & 15.154	 &	  --	    &  --       &         --     &     --       &--\\
\textbf{2076582259160998400} &	     	--          & 4.8721	    & 19.985	 &	  --		&  --       &         --     &     --       &--\\
2076582057294883200 &	5112553\,$^{15L}$ & 5.2427	    & 13.055	 &	5\,812(27)	& 4.595(44)	& -35(4)	& 0.135(25) & 3\\
\textbf{2076396166816178176} &	5200463\,$^{X}$	  & 0.7563	    & 19.301	 &	  --	    &  --       &         --     &     --       &--\\
\textbf{2076581812488757632} &	5112718\,$^{X}$	  & 2.0447	    & 18.446	 &	  --	    &  --       &         --     &     --       &--\\
\textbf{2076581679348337152} &	5112646\,$^{X}$	  & 12.5816	    & 14.484	 &	  --	    &  --       &         --     &     --       &--\\
\textbf{2076394517550928640} &	     	--          & 6.1322	    & 19.448	 &	  --		&  --       &         --     &     --       &--\\
2076394414469300096 &	5113282\,$^{X}$	  & 1.4962	    & 18.000	 &	  --	    &  --       &         --     &     --       &--\\
\textbf{2076394345749739264} &	     	--          & 0.3484	    & 20.151	 & 4\,503(92)	& 2.25(11)& -27.92(44) & 	-0.100(28) & 5 \\
2076394277037190656 &	5113103\,$^{X}$	  & 5.3265	    & 18.427	 &	--		&    --     &      --        &       --      &--\\
\textbf{2076394246980225280} &	5113042\,$^{X}$	  & 12.1961	    & 16.138	 &	--	    &    --     &      --        &       --      &--\\
\textbf{2076394075181511296} &	5112921\,$^{X}$	  & 1.040	    & 15.961	 &	--	    &    --     &      --        &       --      &--\\
\textbf{2076394040821761792} &	5112865\,$^{X}$	  & 8.1032	    & 15.741	 &	--	    &    --     &      --        &       --      &--\\
\textbf{2076394109541233536} &	5112796\,$^{X}$	  & 1.0060	    & 17.176	 &	--	    &    --     &      --        &       --      &--\\
\textbf{2076394105234044032} &	5112746\,$^{X}$	  & 4.5372	    & 18.790	 &	--		&    --     &      --        &       --      &--\\
2076395071614000512 &	5113378\,$^{13L}$ & 24.4924	    & 15.524	 &	--		&    --     &      --        &       --      &--\\
2076394173950795136 &	     	 --         & 3.5243	    & 19.318	 &	--	    &    --     &      --        &       --      &--\\
\textbf{2076393869023082240} &	5113011\,$^{X}$	  & 10.1982	    & 14.427	 &	--	    &    --     &      --        &       --      &--\\
\textbf{2076393834655370240} &	     	 --         & 8.6988	    & 19.625	 &	--		&    --     &      --        &       --      &--\\
\textbf{2076393456700278656} &	     	 --         & 3.7551	    & 20.610	 &	--		&    --     &      --        &       --      &--\\
\textbf{2076393903374713856} &	5112711\,$^{X}$	  & 8.4417	    & 17.731	 &	--	    &    --     &      --        &       --      &--\\
2076393491065999360 &	5113306\,$^{X}$	  & 1.9503	    & 18.317	 &	--	    &    --     &      --        &       --      &--\\
2076393353627026944 &	5113228\,$^{1L}$  & 4.9268	    & 12.332	 & 4\,030(92)	& 1.19(11)& -6.277(22) & -0.100(28) & 5\\
\textbf{2076393795993621376} &	5113037\,$^{X}$	  & 2.6187	    & 20.041	 &     --     &   --      &    --          &    --         &--\\
2076393658557409792 &	5024787\,$^{14L}$ & 12.7227	    & 17.924	 &     --     &   --      &    --          &    --         &--\\
2076392765201345280 &	5025129\,$^{X}$	  & 9.5598	    & 17.830	 &     --     &   --      &    --          &    --         &--\\
\textbf{2076392872590618112} &	     	--          & 7.2291	    & 19.511	 &     --     &   --      &    --          &    --         &--\\
\textbf{2076393250540210432} &	5113521\,$^{X}$	  & 0.9759	    & 18.448	 &     --     &   --      &    --          &    --         &--\\
2076393211878004096 &	5113452\,$^{X}$	  & 0.7334	    & 18.605	 &     --     &   --      &    --          &    --         &--\\
\textbf{2076393113108873344} &	5113418\,$^{X}$	  & 12.8024	    & 17.779	 &     --     &   --      &    --          &    --         &--\\
\textbf{2076393108798746752} &	5113384\,$^{X}$	  & 1.3900	    & 17.050	 &     --     &   --      &    --          &    --         &--\\
\textbf{2076392971359814528} &	5113140\,$^{X}$	  & 7.9415	    & 18.691	 &     --     &   --      &    --          &    --         &--\\
\textbf{2076392838230901632} &	5024960\,$^{X}$	  & 9.2894	    & 15.874	 &     --     &   --      &    --          &    --         &--\\
\textbf{2076392872582614144} &	     	--          & 17.6075	    & 18.606	 &     --     &   --      &    --          &    --         &--\\
\textbf{2076390192531117440} &	5025475\,$^{X}$	  & 5.4480	    & 17.747	 &     --     &   --      &    --          &    --         &--\\
\textbf{2076390055084332672} &	5025381\,$^{X}$	  & 5.8275	    & 19.118	 &     --     &   --      &    --          &    --         &--\\
2076389604105250560 &	5025159\,$^{X}$	  & 3.3906	    & 19.053	 &     --     &   --      &    --          &    --         &--\\
\textbf{2076390020724772736} &	5025605\,$^{X}$	  & 12.1848	    & 18.306	 &     --     &   --      &    --          &    --         &--\\
\textbf{2076389711494726272} &	5025195\,$^{X}$	  & 13.1484	    & 18.311	 &     --     &   --      &    --          &    --         &--\\
\textbf{2076390050781929472} &	5025372\,$^{X}$	  & 0.9905	    & 18.329	 &     --     &   --      &    --          &    --         &--\\
\textbf{2076392559042768896} &	5025092\,$^{X}$	  & 4.4708	    & 19.278	 &     --     &   --      &    --          &    --         &--\\
\textit{\textbf{2076488942409044864}} &	     	--          & 1.4495	    & 	--		 &     --     &   --      &    --          &    --         &--\\
\textit{\textbf{2076581988577868032}} &	     	--          & 3.9648	    & 21.031	 &     --     &   --      &    --          &    --         &--\\
\textit{\textbf{2076488049065148288}} &	5112292\,$^{X}$	  & 1.9006	    & 16.004	 &     --     &   --      &    --          &    --         &--\\
\textit{\textbf{2076487739818146048}} &	     	--          & 3.8497	    & 21.192	 &     --     &   --      &    --          &    --         &--\\
\textit{\textbf{2076487224431395712}} &	     	--          & 2.2195	    & 20.442	 &     --     &   --      &    --          &    --         &--\\
\textit{\textbf{2076581919858187136}} &	     	--          & 7.4127	    & 20.923	 &     --     &   --      &    --          &    --         &--\\
\textit{\textbf{2076581748057139456}} &	     	--          & 17.6544	    & 21.159	 &     --     &   --      &    --          &    --         &--\\
\textbf{2076487464947285248} &	5023707\,$^{X}$	& 0.9582          & 15.199	& 	--		&   --        &     --     &    -- & --      \\
\textit{\textbf{2076392941302191616}} &		     		& 0.6279          & 20.924	&     --      &		--	&	--       & --  & --       \\
\textbf{2076394620627810432} &	5200435\,$^{X}$	& 0.6398          & 18.505	&	--		&   --        &   --       & 	--     & --  \\
\textbf{2076394384411491712} &	5113161\,$^{X}$	& 2.2842          & 19.132	&	--		&     --      &   --       & 	 --    & --  \\
2076491588108524160 &	5111658\,$^{X}$	& 1.3542         & 16.277	& 4\,470(60)	& 4.572(57)	&  -89(1)& -0.76(6) & 2\,$^*$\\
\textbf{2076491588108524160} & -- &  1.9159 & 17.492 & --  & -- & --  & --  & -- \\
\textbf{2076298310287835776} & 4936866\,$^{X}$ & 1.3921 & 18.014 & --  & -- & --  & --  & -- \\
\textbf{2076390123803764352} &  -- & 3.9365 & 15.646 & --  & -- & --  & --  & -- \\
%422-2076392838222892160 &	     	--          & 0	        & 20.602	 &     --     &   --      &    --          &    --         &--\\
\hline\hline
\end{tabular}
}
\end{table*}

%%%%%%%%%%%%%%%%%%%%%%%%%%%%%%%%%%%%%%%%%%%%%%%%%%%%%%%%%%%%%%%%%%%%%
%FIELD UNCLASSIFIED
\begin{table}
\caption{The list of unclassified variables that are not identified as cluster members in our analysis. See captions of Table\,1 and 5 for explanation.}
\label{tab:cluster_unclassified}
\centering
\resizebox{\columnwidth}{!}{
\begin{tabular}{cllc}
\hline\hline
\multicolumn{1}{c}{\multirow{2}{*}{\gaia\ DR3}} & \multicolumn{1}{c}{\multirow{2}{*}{KIC}} & \multicolumn{1}{c}{Period} & \multicolumn{1}{c}{G} \\
&& \multicolumn{1}{c}{[days]} & \multicolumn{1}{c}{[mag]}\\
\hline\hline
\textbf{2076489011128317824} &	5111916\,$^{X}$	& 0.2954  & 19.854 \\
\textbf{2076488495732906496} &	5111771\,$^{X}$	& 0.6336  & 20.106 \\
\textbf{2076487533659553024} &	5112097\,$^{X}$	& 0.2629 & 19.411 \\
\textbf{2076299998215027968} &	--	    		& 0.7388 & 20.198 \\
\textbf{2076487224421838336} &	--	     	    & 0.5368 & 19.785 \\
\textbf{2076299826420547200} &	5024510\,$^{X}$	& 3.1140 & 14.827 \\
\textbf{2076299895135783168} &	--	     		& 0.4290 & 20.698 \\
\textbf{2076298447723494016} &	5024319\,$^{X}$	& 0.3007 & 19.934 \\
\textbf{2076298791320709120} &	5024566\,$^{X}$	& 0.7352 & 18.553 \\
\textbf{2076394517548400896} &	5112901\,$^{X}$	& 0.6911 & 19.291 \\
\textbf{2076394384411456000} &	5113175\,$^{X}$	& 0.1723 & 18.885 \\
\textbf{2076393314960052480} &		--     		& 0.4764 & 20.839 \\
\textbf{2076393108798736384} &	5025311\,$^{X}$	& 1.9564 & 15.472 \\
\textbf{2076392528993235968} &	5024986\,$^{X}$	& 0.3169 & 16.960 \\
\textbf{2076581984282789632} &	5112326\,$^{X}$ & 0.1085 & 17.258 \\
\textbf{2076487838597567104} &	5112332\,$^{X}$	& 6.4325 & 18.557 \\
\textbf{2076393869023083776} & -- & 2.9235 & 18.391  \\
\textbf{2076390123803775104} & -- & 3.3921 & 18.469  \\
\hline
\hline\hline
\end{tabular}
}
\flushleft
\end{table}

%%%%%%%%%%%%%%%%%%%%%%%%%%%%%%%%%%%%%%%%%%%%%%%%%%%%%
%Table of Other non gaia spots
\begin{table}[htbp]
\centering
\caption{A list of equatorial coordinates that are associated with the superstamp pixels showing signal in their amplitude spectra. The only coordinates not marked in bold were associated with a variable star NGC\,6819 SHLP\,25645 and reported by Street \etl(2002).}
\label{tab:coordinates}
\begin{tabular}{ccrc}
\hline\hline
\multicolumn{1}{c}{$\alpha_{\rm 2000}$} & \multicolumn{1}{c}{$\delta_{\rm 2000}$} & \multicolumn{1}{c}{Period} & \multicolumn{1}{c}{\multirow{2}{*}{Type}} \\
\multicolumn{1}{c}{[hh mm ss.ss]} & \multicolumn{1}{c}{[dd mm ss.s]} & \multicolumn{1}{c}{[days]} & \\
\hline\hline
\textit{\textbf{19 41 20.77}}  & \textit{\textbf{+40 17 18.77}}	& 0.8824	    & eclipsing          \\
\textit{\textbf{19 40 57.78}}  & \textit{\textbf{+40 10 11.87}}	& 0.5085	    & active eclipsing \\
\textit{19 40 57.99} & \textit{+40 19 00.30} & 0.2734	    & contact       \\
\textit{\textbf{19 41 0.81 }}    & \textit{\textbf{+40 18 20.35}}	& 0.2342 	    & contact          \\
\textit{\textbf{19 41 10.24}} & \textit{\textbf{+40 10 35.73}} & 0.2080         & contact \\
\textit{\textbf{19 41 23.73}}  & \textit{\textbf{+40 18 54.4}}	& 65.3493	    & outbursting      \\
\textit{\textbf{19 41 28.59}}  & \textit{\textbf{+40 11 31.17}}	& 29.6949	    & outbursting     \\
\textit{\textbf{19 41 29.61}}  & \textit{\textbf{+40 11 19.35}}	& 45.4381	    & outbursting     \\
\textit{\textbf{19 40 48.17}}   & \textit{\textbf{+40 13 12.24}}	& 0.4093	    & rotating         \\
\textit{\textbf{19 41 13.06}}  & \textit{\textbf{+40 10 37.26}}	& 0.6044	    & rotating        \\
\textit{\textbf{19 40 53.45}}  & \textit{\textbf{+40 08 43.40}}	& 0.3261	    & rotating          \\
\textit{\textbf{19 40 57.85}}  & \textit{\textbf{+40 07 31.53}}	& 0.7897	    & rotating           \\
\textit{\textbf{19 41 25.84}}  & \textit{\textbf{+40 08 21.47}}	& 2.6166	    & rotating           \\
\textit{\textbf{19 41 26.11}}  & \textit{\textbf{+40 06 58.64}}	& 1.7384	    & rotating           \\
\textit{\textbf{19 41 15.77}}  & \textit{\textbf{+40 05 52.16}}	& 8.1636	    & rotating           \\
\textit{\textbf{19 41 23.23}}  & \textit{\textbf{+40 18 31.1}}	& 2.2959	    & rotating           \\
\textit{\textbf{19 41 27.12}}  & \textit{\textbf{+40 16 25.39}}	& 3.7529	    & rotating       \\
\textit{\textbf{19 41 16.70}}   & \textit{\textbf{+40 15 34.46}}	& 2.0277	    & rotating           \\
\textit{\textbf{19 41 41.79}}  & \textit{\textbf{+40 16 43.86}}	& 15.2913	    & rotating        \\
\textit{\textbf{19 41 28.35}}  & \textit{\textbf{+40 13 1.84}}	& 41.1247	    & rotating           \\
\textit{\textbf{19 41 47.97}}  & \textit{\textbf{+40 12 21.28}}	& 0.2771	    & rotating        \\
\textit{\textbf{19 41 37.49}}  & \textit{\textbf{+40 09 3.74}}	& 52.5409	    & rotating            \\
\textit{\textbf{19 41 50.89}}  & \textit{\textbf{+40 10 28.79}}	& 0.9870	    & rotating        \\
\textit{\textbf{19 41 6.38}} & \textit{\textbf{+40 14 53.84}} & 0.3660 & unclassified \\
\hline\hline
\end{tabular}
\end{table}
%%%%%%%%%%%%%%%%%%%%%%%%%%%%%%%%%%%%%%%%%%%%%%%%%%%%%%%%%%%%%%%%%%%%%

We also derived distance from the parallaxes of the cluster members with probability membership higher than 90\% and the relative parallax uncertainty smaller than 10\%, which equals 2.48(18)\,kpc. The distance measurements from parallaxes and isochrone fits are in agreement to each other, and the age and [Fe/H] derived in our work is comparable to the results reported by Choi \etl(2018).

We also plotted isochrones in the \teff\ and \logg\ plane (HRD) shown in Fig.\,8. We used the isochrones for the age and [Fe/H], which we derived from the isochrone fitting in the CMD. We overplotted the isochrones with our variable stars from Tables\,1-3, for which \teff\ and \logg\ are listed. The column 'HRD' in Tables\,1-4 describes the location of a given star in the HRD. If the location agrees with the one in the CMD, we can expect the spectroscopic fit is likely correct. We obtained an agreement in all but two cases. One of the exceptions is a rotational variable KIC\,5113295. In both cases, as the CMD shows, the gravity of MS stars is around 4.5.

\begin{figure*}
\centering
\includegraphics[width=0.8\hsize]{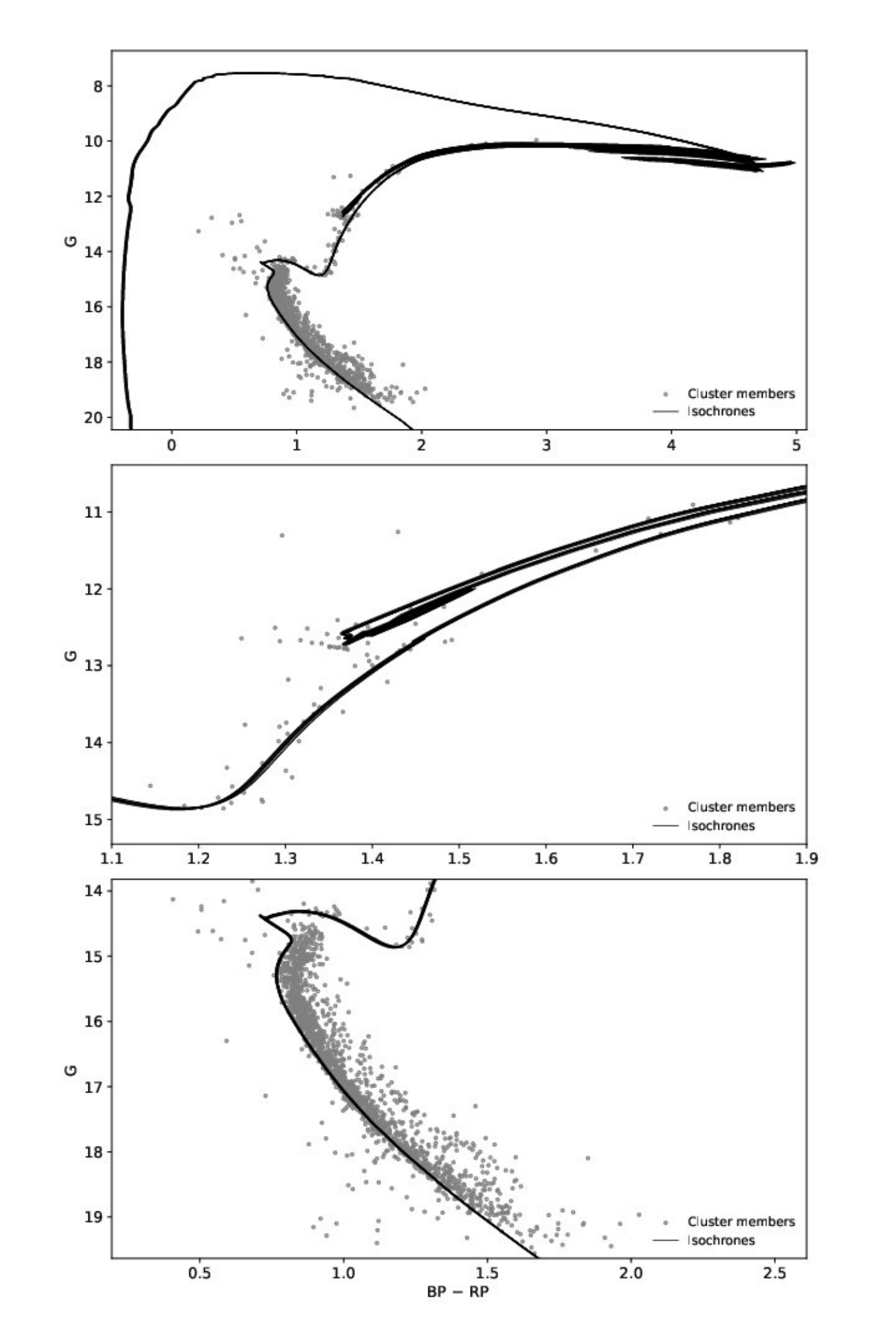}
\caption{Color-magnitude diagram of NGC\,6819, showing the best MIST isochrone fits. The top panel shows the overall diagram, while the middle panel shows the red clump region and the bottom panel shows the main-sequence region.}
\label{isocmd}
\end{figure*}

\begin{figure}[!htb]
\centering
\includegraphics[scale=0.5]{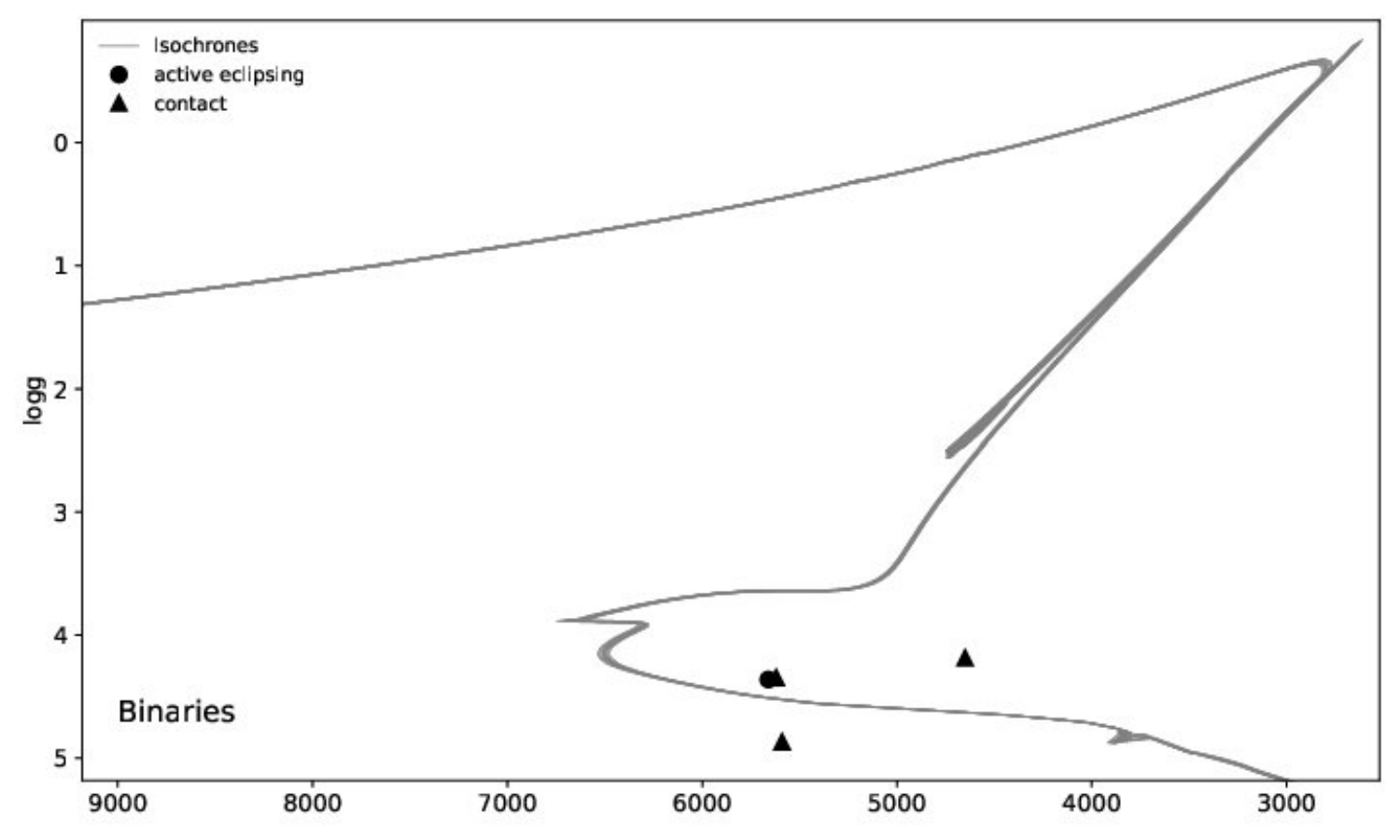} \\
\includegraphics[scale=0.5]{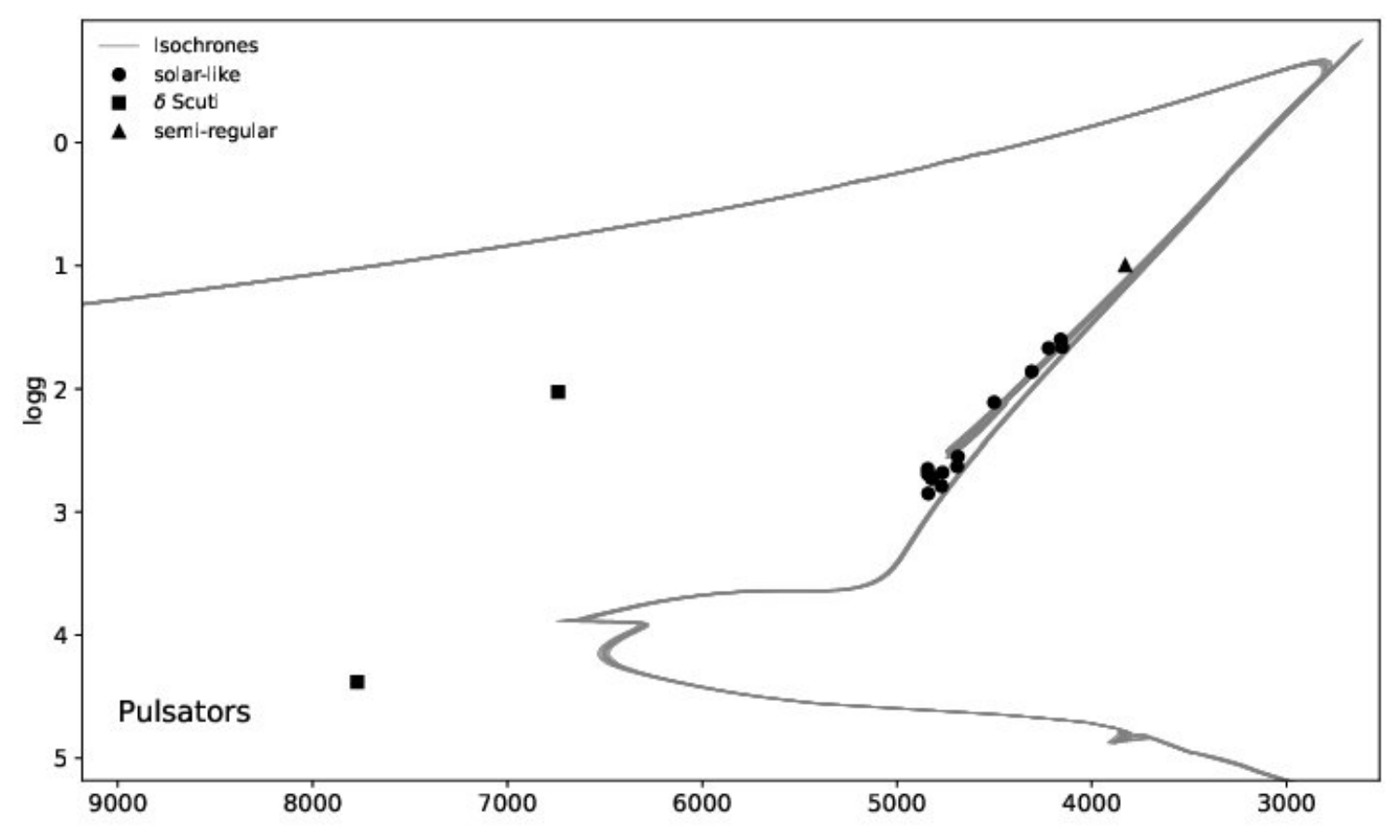} \\
\includegraphics[scale=0.5]{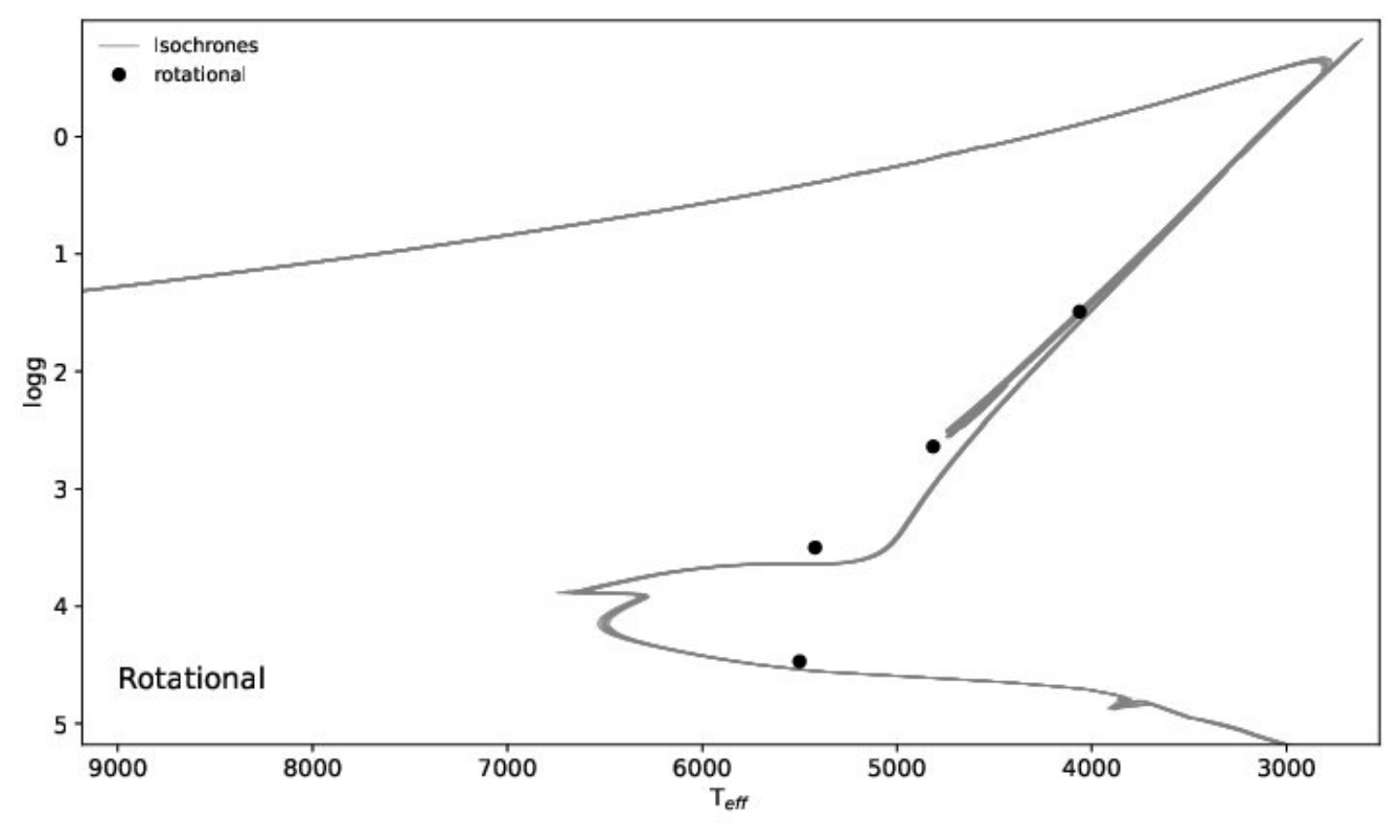}\\
\caption[\teff--\logg\ diagram]{\small \teff--\logg\ diagram showing the cluster members with atmospheric parameters derived in this work. The top, middle, and bottom panels show binary, pulsating, and rotating stars, respectively. The isochrones represent the best MIST fits in the CMD plotted in \teff\ and \logg\ space.}
\label{hrd}
\end{figure}

\section{Summary}
We presented a search for variable stars in the \kep\ superstamp data of NGC\,6819. Individual pixels were searched, by means of a Fourier amplitude spectrum, and contiguous apertures for each object that shows a significant flux variation were defined. The coordinates of these optimal apertures were matched with optical counterparts using the Pan-STARRS survey. In total, we detected 385 variable stars. We searched the literature for variable stars reported prior to our work, and we found that 27 variable stars were reported by Street \etl(2002), eight by Street \etl(2005), one from the All Sky Automated Survey\,(ASAS) catalogue by Pigulski \etl(2009), 75 from VSX by Watson \etl(2006), 21 from General Catalogue of Variable Stars\,(GCVS) and from New Catalogue of Variable Stars\,(NSV) by Samus\,(2017). We found 306 stars having KIC designations, out of which 226 stars do not have any data delivered to the MAST. For 270 stars, their variable nature was unknown prior to our analysis. These stars are marked in bold in Tables\,1-10.
%Just recently, and independently of our work, Colman \etl(2022) reported light curves of KIC stars but did not specify which stars were found to be variables. Accounting for this report, we can consider 71 variable stars having no KIC designation to be new detections of variable stars. These stars have no ground-based data reported.

Using \gaia\,DR3 astrometry, we calculated the membership probabilities for all variable stars in our sample by applying Bayesian Gaussian mixture models. We considered a star to be a cluster member if its membership probability is higher than 50\%. In total, we found 128 cluster member variables, including 17 binaries, 24 pulsators, 82 rotationally and five unclassified. The locations of these cluster variable stars in the CMD diagram indicate their evolutionary status. In the CMD, a majority of our variable stars are located on the MS, while solar-like pulsators are mostly split into the RGB and RC, semi-regular variables are located on the AGB, and five in the BS region.

In the case of selected binary systems, we estimated mid-times of eclipses or minima of a continuous flux variation and derived ephemerides. We calculated the O-C diagrams and checked for any orbital period variation. Four binary systems show significant period variation, however its nature is not confirmed. The solar-like as well as $\delta$\,Scuti and $\gamma$\,Doradus pulsators have been a subject of a detailed analysis of its pulsation content, and the results will be published by Themessl \etl (in prep.) and Guzik \etl(2023), respectively.

We used publicly available spectra of 42 variable stars. In addition, we analyzed spectra of eight stars collected with the 2.51\,m NOT and spectra of five stars were taken with the 3.5\,m APO telescopes. For the latter spectra, we quote a minimum uncertainty of 200\,K due to the low S/N. Spectra of 21 stars were fitted with the {\sc XTgrid}, while for the remainder of the sample, we adopted the fit values from the surveys. We derived \teff, \logg, [Fe/H] and RVs. Thanks to our spectral analysis we were able to recover consistent stellar parameters from very diverse spectroscopic data. This consistency is reflected by the similar distribution of stars in the CMD and HRD.

We used MIST isochrones to fit the positions of the cluster members in the CMD. Our  variable star population was also included in the fit. The best solution was achieved for six isochrones giving [Fe/H] of -0.01(2) and the age of 2.54(3)\,Gyr. The age estimate agrees with the value reported by Choi \etl(2018). The average distance estimate from the distance modulus is 2.3\,kpc, which is in agreement with our independent estimate of 2.48(18)\,kpc derived from the \gaia\ astrometry of a selected sample of the cluster members.

\section*{Acknowledgement}
Financial support from the National Science Centre under projects No.\,UMO-2017/26/E/ST9/00703 and UMO-2017/25/B/ST9/02218 is acknowledged. PN acknowledges support from the Grant Agency of the Czech Republic (GA\v{C}R 22-34467S). The Astronomical Institute in Ond\v{r}ejov is supported by the project RVO:67985815. This paper includes data collected by the Kepler mission and obtained from the MAST data archive at the Space Telescope Science Institute (STScI). Funding for the Kepler mission is provided by the NASA Science Mission Directorate. STScI is operated by the Association of Universities for Research in Astronomy, Inc., under NASA contract NAS 5-26555. This work has made use of data from the European Space Agency (ESA) mission. \gaia\ (https://www.cosmos.esa.int/gaia), processed by the \gaia\ Data Processing and Analysis Consortium (https://www.cosmos.esa.int/web/gaia/dpac/consortium). Funding for the DPAC has been provided by national institutions, in particular, the institutions participating in the \gaia\ Multilateral Agreement. This research has made use of the NASA/IPAC Extragalactic Database (NED), which is operated by the Jet Propulsion Laboratory, California Institute of Technology, under contract with the National Aeronautics and Space Administration. Based on observations obtained with the Apache Point Observatory 3.5-meter telescope, which is owned and operated by the Astrophysical Research Consortium  (https://www.apo.nmsu.edu). This research has used the services of \url{www.Astroserver.org} under reference YIE7AQ and ZLNV9K.
%We thank the anonymous referees for valuable comments, which have significantly improved the quality of the manuscript.
%{\bf We acknowledge that we have used the following publicly archived spectras, from Large Sky Area Multi-Object Fibre Spectroscopic Telescope (LAMOST), which is a National Major Scientific Project built by the Chinese Academy of Sciences\,(http://dr6.lamost.org), APO Galactic Evolution Experiment (APOGEE) in New Mexico, United States\,(https://www.sdss.org/dr16/irspec), Hectospec 300 optical fiber fed spectrograph commissioned at the MMT in the spring of 2004\,(https://oirsa.cfa.harvard.edu).}

\section*{References}
\small
%%%%%
Ahn C. P. et al., 2014, ApJ, 211, 17 \\
Ak T., Bostanci Z. F., Yontan T., Bilir S., Guver T., Ak S., Urgup H., Paun-zen E., 2016, Ap\&SS, 361, 126 \\
Anthony-Twarog B. J., Deliyannis C. P., Twarog B. A., 2014, AJ, 148, 51 \\
Auner G., 1974, A\&AS, 13, 143 \\
Balona L. A., et al., 2013, MNRAS, 430, 3472 \\
Baran A., 2013, AcA, 63, 203 \\
Barkhatova K.A., Vasilevsky A.E., 1967, Peremennye Zvezdy, 16, 191 \\
Barnard E. E., 1931, Publications of the Yerkes Observatory, 6, 1 \\
% Uncomment the line below to include the reference
%Barnes S.A., 2003, ApJ, 586, 464-479 \\
Basu S. et al., 2011, ApJ, 729, L10 \\
Bohlin, R.~C., M{\'e}sz{\'a}ros, S., Fleming, S.~W., et al., 2017, AJ, 153, 234 \\
Bragaglia A., et al., 2001, AJ, 121, 327 \\
Burkhead M. S., 1971, AJ, 76, 251 \\
Cantat-Gaudin T., et al, 2018, AAP, 618, A93 \\
Chambers K. C. et al., 2016, arXiv, 1612, 05560 \\
Choi J., Dotter A., Conroy C., Cantiello M., Paxton B., Johnson B. D., 2016, ApJ, 823, 102 \\
Choi J., et al., 2018, ApJ, 863, 65 \\
Colman I. L. et al., 2022, ApJS, 258, 39 \\
Dotter A., 2016, ApJS, 222, 8 \\
Fabricant D. et al., 2005, PASP, 117, 1411 \\
Ferguson T. S., 1973, The Annals of Statistics, 1, 209 \\
Flewelling H. A. et al., 2020, ApJ, 251, 7 \\
Friel, et al., 1989, PASP, 101, 1105-1112 \\
% Uncomment the line below to include the reference
%Gaia Collaboration et al., 2016, A\&A, 595, A1 \\
Gaia Collaboration et al., 2022, A\&A, accepted \\
Gao X.-H., Xu S.-K., Chen L., 2015, Research in Astronomy and Astrophysics, 15, 2193 \\
Gosnell N. M., et al., 2012, APJ, 745, 57 \\
Guzik J., et al., 2023, ApJ, accepted \\
Hole K. T., Geller A. M., Mathieu R. D., Platais I., Meibom S., Latham D. W., 2009, AJ, 138, 159 \\
Hoskin M., 2005, Journal for the History of Astronomy, 36, 373-406 \\
Kalirai J. S., et al., 2001, AJ, 122, 266 \\
Kaluzny J., Shara M. M., 1988, AJ, 95, 785 \\
Kamann S., Bastian N. J., Gieles M., Balbinot E., Hénault-Brunet V., 2018, MNRAS, 483, 2197 \\
Kinemuchi K., et al., 2012, Publications of the Astronomical Society of the Pacific, 124, 963 \\
King I. R., 1964, Royal Greenwich Observatory Bulletins, 82, 106 \\
Kryachko T.V., 2001, Information Bulletin on Variable Stars, 5058, 1 \\
Kwee K., van Woerden H., 1956, Bulletin of the Astronomical Institutes of the Netherlands, 12, 327 \\
Lee-Brown D. B., Anthony-Twarog B. J., Deliyannis C. P., Rich E., Twarog B. A., 2015, AJ, 149, 121 \\
Lindoff U., 1972, A\&AS, 7, 497 \\
Lindoff U., 1971, Information Bulletin on Variable Stars, 606, 01 \\
Majewski S. R, et al., 2017, AJ, 154, 94 \\
Manteiga M., et al., 1989, AAP, 210, 66-77 \\
N{\'e}meth, P., Kawka, A., Vennes, S., 2012, MNRAS, 427, 2180 \\
Pedregosa F., et al., 2011, Journal of Machine Learning Research, 12, 2825 \\
Pigulski A., et al., 2009, AcA, 59, 33 \\
Platais I., Gosnell N. M., Meibom S., Kozhurina-Platais V., Bellini A., Veillet C., Burkhead M. S., 2013, AJ, 146, 43 \\
Purgathofer A., 1966, Mitteilungen der Universitaets-Sternwarte Wien, 13, 7 \\
Rosvick J. M., Vandenberg D. A., 1998, AJ, 115, 1516 \\
Salaris M., Weiss A., Percival S. M., 2004, AAP, 414, 163 \\
Sampedro L., et al., 2017, MNRAS, 470, 3937 \\
Samus N.N., et al., 2017, Astronomy Reports, 61, 80-88 \\
Sanders W. L., 1972, A\&A, 19, 155 \\
Samus N.N., et al., 2017, Astronomy Reports, 61, 80-88 \\
Sanders W. L., 1972, A\&A, 19, 155 \\
Sanjayan S., et al., 2022a, AcA, 72, 77 \\
Sanjayan S., et al., 2022b, MNRAS, 509, 763-777 \\
%Skumanich A., 1972, ApJ, 171, 565 \\
Stello D., et al., 2010, Astronomische Nachrichten, 331, 985 \\
Stello D., et al., 2011, ApJ, 739, 13 \\
Street R. A., et al., 2002, MNRAS, 330, 737 \\
Street R. A., et al., 2003, MNRAS, 340, 1287 \\
Street R. A., et al., 2005, MNRAS, 358, 795 \\
Talamantes  A.,  Sandquist  E.  L.,  Clem  J.  L.,  Robb  R.  M.,  Balam  D.  D.,Shetrone M., 2010, AJ, 140, 1268 \\
Themessl N., et al., in prep., -, -, - \\
Tody D., 1986, SPIE Conference Series, 627, 733 \\
Tody D., 1993, ASP  Conference Series, 52, 173 \\
Watson C.L.,Henden A.A. and Price A., Society for Astronomical Sciences Annual Symposium, 2006, 25, 47 \\
%Yang S. C., Sarajedini A., Deliyannis C. P., Sarrazine A., 2012, American Astronomical Society Meeting, 22, 15 \\
Zhang Bo., et al., 2015, Research in Astronomy and Astrophysics, 15, 1197 \\
Zhao G.et al., 2012, RA\&A, 12, 7

\end{document}